\documentclass[12pt,peerreview,draftclsnofoot,onecolumn]{IEEEtran}

\usepackage{amsmath,amssymb}
\usepackage{cite}
\usepackage{graphicx}
\usepackage{epstopdf}
\usepackage{xcolor}
\usepackage{cases}
\usepackage{url}

  \newtheorem{proposition}{Proposition}
 \newtheorem{corollary}{Corollary}
 
  \newtheorem{remark}{Remark}

\begin{document}
\title{Adaptive Deployment for UAV-Aided Communication Networks}

\author{Zhe~Wang,~\IEEEmembership{Member,~IEEE}, Lingjie Duan,~\IEEEmembership{Senior Member,~IEEE},\\ and Rui Zhang,~\IEEEmembership{Fellow,~IEEE}
\thanks{Part of this work was presented at the IEEE Global Communications Conference (GLOBECOM 2018), Abu Dhabi, UAE, Dec. 9-13, 2018.~\cite{Zhe}}
\thanks{Z. Wang and L. Duan are with Pillar of Engineering Systems and Design, Singapore University of Technology and Design,
Singapore (e-mail: zhe$\_$wang@sutd.edu.sg; lingjie$\_$duan@sutd.edu.sg).
}
\thanks{R. Zhang is with the Department of Electrical and Computer Engineering,
National University of Singapore, Singapore (e-mail:
elezhang@nus.edu.sg).
}
}

\maketitle

\begin{abstract}

Unmanned aerial vehicle (UAV) as an aerial base station is a promising technology to rapidly provide  wireless connectivity to ground users. Given UAV's agility and mobility, a key question is how to adapt UAV deployment to best cater to the instantaneous wireless traffic in a territory. In this paper, we propose an adaptive deployment scheme for  a UAV-aided communication network, where the UAV adapts its displacement direction and distance to serve randomly moving users' instantaneous traffic in the target cell.
In our adaptive scheme, the UAV does not need to learn users' exact locations in real time, but chooses its displacement direction based on a simple majority rule by flying to the spatial sector
with the greatest number of users in the cell. To balance the service qualities of the users in different sectors, we further optimize the UAV's displacement distance in the chosen sector to maximize the average throughput and the successful transmission probability, respectively.
We prove that the optimal displacement distance for average throughput maximization decreases with the user density: the UAV moves to the center of the chosen sector when the user density is small and the UAV displacement becomes mild when the user density is large.
In contrast, the optimal displacement distance for success probability maximization does not necessarily decrease with the user density and further depends on the target signal-to-noise ratio (SNR) threshold.
Extensive simulations show that the proposed adaptive deployment scheme
outperforms the traditional non-adaptive scheme, especially when the user density is not large.

\end{abstract}


\section{Introduction}

Unmanned aerial vehicles (UAVs) aided wireless communication network is  a promising solution to rapidly establish wireless connectivity for the mobile devices that are beyond the coverage of the terrestrial communication infrastructure
\cite{UAV}.
For example, UAVs can work as mobile aerial base stations to provide emergency communication services to the ground terminals in battle fields, disaster scenes, congested roads, blind spots and rural areas.
UAV-aided communication has two major advantages. First, with the agility and mobility features, a UAV can rapidly fly to serve the users closely and adapt to the on-demand surge. Second, compared with the traditional terrestrial base stations, the UAV operating at a high altitude connects to ground users via more reliable communication channels thanks to line-of-sight (LoS) links.

A key design challenge of UAV-aided communication network is how to deploy a UAV to cater to wireless users' instantaneous traffic demands.
Recent works have studied the UAV deployment for the static user networks, where the ground users' locations are fixed and known.
For example, \cite{YZeng0,LSong,YZeng1,Lyu2,Simeone} design the trajectory of a single UAV to relay information \cite{YZeng0,LSong}, broadcast/multicast messages \cite{YZeng1},  or provide offloading services to the ground users \cite{Lyu2,Simeone}, in the delay-tolerant systems. In \cite{QWu}, cooperative trajectory design for multiple UAVs is further investigated to mitigate the mutual co-channel interference.
Besides trajectory designs, researchers investigate the UAV deployment that
provides the wireless coverage to the static users in a target geographical area, by designing the optimal operating location in three-dimensional (3D) airspace \cite{MM0,Akram,coverage}, minimizing the total deployment time \cite{Xiao}, or minimizing the number of the stop points for the UAV \cite{MM}. \cite{Xinping} and \cite{Xuehe} study the economic issues, i.e., mechanism design and dynamic service pricing, in the multi-user UAV-aided network.
In \cite{Juting2}, the authors propose a machine learning approach to reconstruct a radio map of the air-to-ground channel across a dense urban environment, which is then exploited to search the global optimal UAV positioning for establishing the best wireless relay link between a BS and a static user in \cite{Juting1}.

Compared with static user networks, the UAV deployment design for random user networks  is more challenging when the user locations are random. \cite{MChen} proposes a machine learning approach to predict the user behaviors (i.e., content request distribution and mobility pattern) and designs the UAV deployment to meet the users' quality of experience requirement while minimizing the UAV's transmit power. In some other scenarios, the users' locations are highly random and unpredictable. Recent works in \cite{JGuo,Harp,BG,Chiya,Ultra} adopt Poisson point process (PPP) to analytically characterize the spatial randomness of the users in the UAV-aided wireless networks. In \cite{JGuo,Harp,BG}, the uplink or downlink  coverage probability  of the UAV is analyzed  by considering the user distribution follows PPP, where the effect of users' location randomness is captured from a long-term average perspective. \cite{Chiya} derives the optimal density of aerial base stations in a spectrum sharing scenario to maximize the drone small-cell network throughput while satisfying the cellular
network efficiency constraint.
\cite{Ultra} derives the UAV's optimal altitude that maximizes the coverage region by guaranteeing a minimum outage performance over the region. However, in the aforementioned literature, the UAVs are deployed in a probabilistic or average sense, i.e., the UAVs are either statically located in the cell center \cite{JGuo}\cite{Ultra} or randomly located following a Binomial/Poisson point process \cite{Harp,BG,Chiya},
where the specific UAVs' locations in each time/realization are independent of the locations of the nearby users. This non-adaptive UAV deployment cannot cater well to the real-time demands of the mobile users.

To solve this problem, we study the adaptive UAV deployment that allows the UAV to adapt its location in each realization to the dynamic locations of the nearby users. We model the users to follow a homogeneous PPP, where the number of users in the target cell is a Poisson random variable that changes across different time realizations. In this type of random network, the ground base station is traditionally deployed at the center of each cell due to the uniform user density \cite{JGuo}\cite{Ultra}.
However, this deployment strategy may not be efficient since the instantaneous user traffic load can be asymmetric in the cell, e.g., there are more users in some hot-spot areas than others. To fully exploit the mobility feature of the UAV, it would be desirable to deploy the UAV adaptively according to the users' spatial and temporal demand changes. Furthermore, in practice, it can be difficult for the UAV to precisely know the exact locations of the mobile users upon deployment, especially when the nearby terrestrial base stations experience congestion or failure for helping user localization \cite{Xinping}. The UAV may only have limited side information of user locations in real time, e.g., estimated user number in each service area/sector of the target cell. In this work, we propose a traffic-aware adaptive UAV deployment scheme in UAV-enabled communication networks, where the UAV serves a random number of Poisson distributed mobile users in its cell and makes deployment decisions according to the asymmetric  instantaneous traffic in all sectors of the cell. Unlike the previous works that consider the non-adaptive UAV deployment with the Poisson distributed users in \cite{JGuo,Harp,Chiya,BG,Ultra}, our study characterizes explicitly the correlation between the users' instantaneous locations and the UAV's displaced location to serve such users. To the best of our knowledge, this is the first analytical work that studies the traffic-aware adaptive UAV deployment under the limited side information of user locations. The key contributions of this work are summarized as follows.
\begin{itemize}
\item \emph{Novel traffic-aware adaptive UAV deployment:}
To best cater to the instantaneous user traffic in each realization, we propose a traffic-aware adaptive UAV deployment scheme in Section~II, where the UAV only needs to know the numbers of users in different spatial sectors and decides where to fly by following a simple majority-vote rule regarding  users' numbers in different sectors. That is, the UAV flies to the sector that has the greatest number of users with a certain displacement distance within the sector, where the optimal distance is further designed by maximizing different quality of service (QoS) objectives.
\item \emph{Average throughput maximization via optimal UAV displacement:}
     In Section III, we consider a delay-tolerant variable-rate system, where the UAV (or user) transmits signal in downlink (or uplink) with best-effort by adapting its rate according to the transmission distance. For this case, we derive the average throughput of the users in a one-dimensional (1D) ground network under the proposed adaptive deployment scheme. The optimal displacement distance in the chosen sector is further designed by maximizing the average throughput of the users in the cell.
     We show that the optimal displacement distance decreases with the user density, i.e., the UAV should move to the center of the intended sector  in the low user density regime and the displacement distance is small when the user density is large.
\item \emph{Successful transmission probability maximization via optimal UAV displacement:}
    In Section~IV, we consider a delay-limited fixed-rate system, where the UAV (or user) transmits with a fixed rate and a transmission is successful if the signal-to-noise ratio (SNR) at the receiver exceeds a target threshold. We derive the optimal displacement distance in the chosen sector to maximize the success  probability of an arbitrary user in the cell. We show that the optimal displacement distance critically depends on the target SNR. In the high target SNR regime, the UAV's coverage region is not large enough to cover the intended sector. As a result, the corresponding optimal UAV placement is at any point ensuring that its coverage region is within the chosen sector.
    While in the low target SNR regime,
    the UAV is able to serve not only the users in the intended sector but also those in the neighboring sector. The corresponding optimal displacement distance thus becomes  unique and increases with the SNR threshold.
\item \emph{Extension to a 2D ground network:}
   We further extend the deployment design and analysis to  a 2D random user network in Section V, where the UAV has multiple displacement directions to adapt in general.
   We consider the similar user-number based  majority-vote rule for choosing the UAV displacement sector and further optimize the UAV displacement distance under the objectives of average throughput maximization and success probability maximization, respectively. We show that most of the main results for the 1D network hold for the 2D scenario.  A key difference is that, the optimal displacement distance that maximizes the successful transmission probability changes  with the user density in the low target SNR regime for the 2D network, while it is independent of the user density for the 1D network.
   Finally, we show that the proposed adaptive deployment scheme outperforms the traditional non-adaptive scheme in terms of both average throughput and success probability in both the 1D and 2D scenarios, especially when the user density is not large. Moreover, in Section VI, we numerically show that the performance can be further improved if the UAV ideally has precise information of user locations in real time.
\end{itemize}

\section{System Model}

\begin{figure}[t!]
    \begin{center}
        \includegraphics[width=0.43\columnwidth]{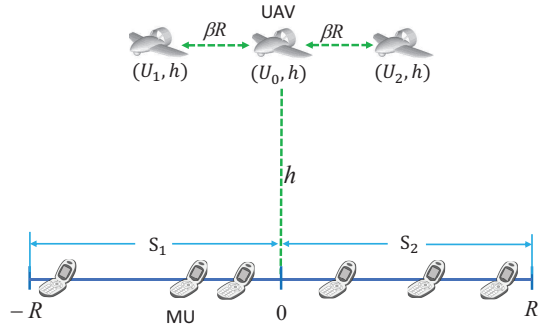}
        \caption{A UAV-aided 1D wireless network, where the MUs are distributed randomly inside the target cell in the range of $[-R, R]$.
         In each realization, the UAV
        chooses one of the displacement positions
        $(U_0,h)=(0,h)$, $(U_1,h)=(-\beta{R},h)$ or $(U_2,h)=(\beta{R},h)$.        }
    \end{center}
\end{figure}

To start with, we consider a 1D  terrestrial user network shown in Fig.~1, where a UAV serves a group of mobile users (MUs) within its cell boundary in either uplink or downlink.
Each MU is allocated with a fixed bandwidth, where we assume the number of channels is always sufficient (e.g., narrowband  FDMA systems \cite{Goldsmithbook}).
The transmit power of the UAV in the downlink (or user in the uplink) is $P_t$.
The MUs (e.g., on an avenue) follow a 1D homogeneous PPP $\{X_m\}$ with density $\lambda$, where $X_m$ is the coordinate of MU$_m$ on the ground. One application of such 1D scenario is to consider that the UAV is providing coverage to a road passing through a rural area.
Later in Section V, we will extend our deployment scheme and analysis to a 2D scenario.
We assume that the UAV is operating at a fixed minimum altitude $h$ under the air traffic control.
The initial location of the UAV is  at the center of the target cell, i.e., $(U_0,h)=(0,h)$.
The target service area of the UAV on the ground is a line segment $S=[-R,R]$, which is partitioned into the left spatial sector $S_1=[-R,0)$ and right sector $S_2=[0,R]$. In one realization of the network, we denote the numbers of the MUs inside $S_1$ and $S_2$ as $k_1$ and $k_2$, respectively, and denote the total number of the MUs inside the cell as $k=k_1+k_2$. 

We now propose a traffic-aware adaptive UAV deployment scheme, where the UAV does not know users' exact instantaneous locations due to the lack of precise user positioning technique in practical situations (e.g. search and rescue). Instead, it relocates according to the side information on user locations, i.e., the numbers $k_1$ and $k_2$, which may change over time.
In each realization, the UAV chooses one out of the three deployment positions $(U_j,h)$ (for $j=0,1,2$) in Fig. 1 as its new displacement location by following a simple majority-vote rule, i.e.,
\begin{subnumcases}
{U=}
U_0=0,~~~~~\text{if $k_1=k_2$} \label{Y2}\\
U_1=-\beta{R},~\text{if $k_1>{k_2}$} \label{Y1}\\
U_2=\beta{R},~~~\text{if ${k}_1<{k_2}$}\label{Y3}.
\end{subnumcases}
We denote $\beta\in[0,1]$ as the displacement factor to reach the displacement distance $\beta{R}$ in each sector,
which will be designed optimally for maximizing the average throughput in Sections~III and maximizing the success probability in Section~IV, respectively. The UAV keeps updating its displacement location $U_j$ according to (1) across different network realizations.\footnotemark
\footnotetext{We consider the UAV has fast moving capability and the user numbers change in a much slower pace as compared with the relocation time of the UAV.}
There is a natural tradeoff in the design of $\beta$.
If the UAV moves into one sector, it shortens the distance from the cell-edge users in this sector, which however comes at the cost of farther distance from the users located opposite to its moving direction in this sector and all users in the other sector. We aim at designing the optimal $\beta$ that maximizes the performance of the overall network to give a good balance between the two sides.
For ease of implementation, we adopt the same
displacement factor $\beta$ across the realizations and design the optimal $\beta^*$ offline by maximizing the long-term average performance per user.
Once the optimal $\beta^*$ is obtained, the UAV chooses its displacement direction according to (1) for each realization in the online operation.
In Section~VI, we will show that updating with a different $\beta$ in each realization
does not bring in significant gain in a long run.
Furthermore, we consider our UAV adaptation scheme is not affected by the lifetime issue of the UAV since the adaptation time of the UAV (e.g., seconds or minutes) is usually much
shorter compared with its total operation time (e.g., up to a few hours).\footnotemark
\footnotetext{
Moreover, the energy sustainability issue can be solved by periodically replacing the battery or by considering a tethered UAV that is connected with a moving ground vehicle which provides a continuous power supply to the UAV through the hardwire tether from the ground.}

For performance evaluation, we randomly select a typical user, i.e., MU$_0$, from the $k$ users (assuming $k>0$) in each realization. Since the user locations are independently and identical distributed (i.i.d.),
the analysis of the average performance of the typical user is equivalent to that of any arbitrary user in the random network.
To model the air-to-ground channels between the UAV and the users, we choose Rician fading which consists of a line-of-sight (LoS) component and a large number of i.i.d. reflected and scattered waves. Under Rician fading, the received signal amplitude
$y$ follows a Rician distribution with the probability density function (PDF) of
\begin{align}
p_Y(y)=\frac{2y(\kappa+1)}{P_r}\exp\left[-\kappa-\frac{(\kappa+1)y^2}{P_r}\right]I_0\left(2y\sqrt{\frac{\kappa(\kappa+1)}{P_r}}\right),~y\geq0,
\end{align}
where $P_r$ is the average received power at the UAV (or MU), $\kappa$ factor is the ratio between the energy in the LoS component and the energy in the NLoS component, and $I_0{(.)}$ is the modified Bessel function of the first kind and zeroth order \cite{Goldsmithbook}.
In the sub-urban and rural areas, LoS link is usually dominating any other links \cite{Harp}.
This corresponds to Rician fading with high $\kappa$ factor.
In Sections III-V, we will focus on analyzing the long-term average user performance from an ergodic sense, which means to average over the short-term fading channels caused by the NLoS components. The direct analysis is difficult due to the existence of Bessel function in the Rician PDF. However, by using Jensen's inequality, one can deduce that the ergordic performance metrics are tightly upper-bounded (approximated) by that of LoS link for high $\kappa$ factor. In other words, the LoS channel can provide a good approximation for the air-ground channel in suburban scenarios.
As such, similar to [3,5,6,8], we model
the communication channel between the UAV and each MU$_0$ by the LoS link that follows the free-space path loss model \cite[Chapter 2]{Goldsmithbook}, i.e.,
\begin{align}
g_{0}=\theta\left[{\sqrt{|X_0-U|^2+h^2}}/{d}\right]^{-2},
\end{align}
where $\theta(dB)=-20\log_{10}\left({4\pi{d}}/{\nu}\right)$ denotes the channel power at the reference distance of $d$ with wavelength $\nu$. We adopt $d=1$ meter throughout the paper and assume the additive white Gaussian noise has zero mean and variance $\sigma^2$.
Hence, the signal-to-noise ratio (SNR) of MU$_0$ is given by
\begin{align}
\gamma=\frac{P_tg_0}{\sigma^2}
=\frac{P_t\theta}{\left(|X_0-U|^2+h^2\right)\sigma^2}.\label{gamma}
\end{align}
For simplicity, we denote $a=P_t\theta/\sigma^2$ in the rest of the paper.

In the following two sections, we will design the optimal displacement factor $\beta$ that maximizes the average throughput and success probability of the typical user in the 1D scenario, respectively.

\section{Average Throughput Maximization With 1D Adaptive UAV Deployment}

In this section, we study the design of the 1D displacement distance $\beta{R}$ for average throughput maximization under the adaptive deployment scheme in (1). For users' traffic, we consider  the  best-effort data applications such as web browsing and video streaming, which are in general delay-tolerant  and admit variable-rate transmission.
To quantify how the traffic load affects the design of $\beta$ from an average perspective, we adopt the same $\beta$ across different MU number and location realizations that maximizes the long-term average throughput of MUs. Note that here we decide the adaptive UAV deployment according to the relationship between $k_1$ and $k_2$ as given in (1). In Appendix A, we further extend our study to design a different $\beta$ for each realization according to the exact numbers of $k_1$ and $k_2$, and the performance improvement of this scheme over that given in (1) is shown to be very  mild in Section VI.

Given the specific locations of the UAV and the typical user MU$_0$ in a realization, the throughput of MU$_0$ is\footnotemark
\footnotetext{Here we consider base 2 for the logarithm as the throughput is measured in bits/second/Hertz (bps/Hz).}
\begin{align}
C=\log(1+\gamma).\label{C0}
\end{align}
To evaluate \eqref{C0} for our adaptive UAV deployment scheme,
we should be able to
find at least one typical user ($k\geq{1}$) inside the cell by excluding the no user case. Otherwise, it does not matter how the UAV moves given zero user to serve.
By taking the expectation of \eqref{C0} over all three location candidates of the UAV, i.e., $U_j$ ($j=0,1,2$) and the two sectors that MU$_0$ may belong to, i.e., $X_0\in{S_i}$ ($i=1,2$),
the average throughput of MU$_0$ is given by
\begin{align}
\mathbb{E}[C\big|k\geq{1}]=\mathbb{E}\left[\log(1+\gamma)\big|k\geq{1}\right]=
\sum\limits_{j=0}^{2}\sum\limits_{i=1}^{2}\omega_{i,j}q_{i,j},\label{ECCC}
\end{align}
where we denote the joint probability that UAV is displaced to $U_j$ and MU$_0$ is inside $S_i$ as
\begin{align}
q_{i,j}=\Pr\left(X_0\in{S_i},U=U_j\big|k\geq{1}\right)\label{qijij},
\end{align}
and the conditional average throughput of the MU$_0$ given it is inside the sector $S_i$ and UAV is displaced to $U_j$ as
\begin{align}
\omega_{i,j}=\mathbb{E}\left[\log(1+\gamma)\big|X_0\in{S_i}, U=U_j,k\geq{1}\right].\label{omegaij}
\end{align}
Since the analysis is symmetric for the cases of $X_0\in{S_1}$ and $X_0\in{S_2}$, we can replace $\sum_{i=1}^{2}$ in \eqref{ECCC} by $2$ and just focus on the analysis of the case of $X_0\in{S_1}$. As such, we rewrite \eqref{ECCC} as
\begin{align}
\mathbb{E}[C\big|k\geq{1}]&=2\sum\limits_{j=0}^{2}q_{1,j}\omega_{1,j}.\label{Exp}
\end{align}
By a slightly abuse of notation, we replace $q_{1,j}$ and $\omega_{1,j}$ by $q_{j}$ and $\omega_j$ in the rest of the paper.

\subsection{Analysis of UAV Displacement Probability $q_j$}
We first derive the joint probability of $q_j$ in \eqref{qijij}. Note that the events of $U=U_j$ and $X_0\in{S_i}$ are correlated since MU$_0$ is one of the $k$ users whose locations affect the displacement decision of the UAV.
According to the probability chain rule, we have
\begin{align}
q_{j}
&=\sum\limits_{K=1}^{\infty}\Pr\left(k=K\big|k\geq{1}\right)\Pr\left(X_0\in{S_1}\big|k=K,k\geq{1}\right)\nonumber\\
&~~~~~~\times\Pr\left(U=U_j\big|X_0\in{S_1},k=K,k\geq{1}\right).\label{qqjj}
\end{align}
By further derivations, we obtain $q_j$ in the following proposition.
\begin{proposition}
The joint probability that the UAV is displaced to $(U_j,h)$ while $X_0\in{S_1}$ is
\begin{subnumcases}{q_j=\nonumber}
\frac{1}{2}\sum\limits_{K=1}^{\infty}\frac{e^{-\mu}{\mu}^{K}}{K!\left(1-e^{-\mu}\right)}\sum\limits_{n=0}^{\lfloor{\frac{K-1}{2}\rfloor}}C_{K-1}^{n}\left(\frac{1}{2}\right)^{n}\left(\frac{1}{2}\right)^{K-n-1},~~~~\text{for}~j=1\label{qj1}\\
\frac{1}{2}\sum\limits_{K=1}^{\infty}\frac{e^{-\mu}{\mu}^{K}}{K!\left(1-e^{-\mu}\right)}
\sum\limits_{n=0}^{\lfloor{\frac{K-1}{2}\rfloor}-1}C_{K-1}^{n}\left(\frac{1}{2}\right)^{n}\left(\frac{1}{2}\right)^{K-n-1},~\text{for}~j=2\label{qj2}\\
\frac{1}{2}-q_1-q_2,~~~~~~~~~~~~~~~~~~~~~~~~~~~~~~~~~~~~~~~~~~~~~~~~~\text{for}~j=0.\label{qj3}
\end{subnumcases}
\end{proposition}
\begin{IEEEproof}
First, we derive the first term $\Pr\left(k=K\big|k\geq{1}\right)$
in the summation of \eqref{qqjj}. Since the MUs follow PPP with density of $\lambda$, the total user number $k$ inside the target cell $[-R,R]$ is a Poisson random variable with the mean value of
$\mu=2\lambda{R}$. We denote $\mu$ as the average traffic load in the cell.
The  probability mass function (PMF) is given by
$\Pr\left(k=K\right)=\frac{e^{-\mu}{\mu}^{K}}{K!}$, denoted by $\text{Poi}(\mu)$.
Conditioned on $k\geq{1}$, the conditional PMF  is thus given by
\begin{subnumcases}
{\Pr\left(k=K\big|k\geq{1}\right)=}
0,~~~~~~~~~~~~~~~\text{for $K=0$}, \label{ConK0} \\
\frac{e^{-\mu}{\mu}^{K}}{K!\left(1-e^{-\mu}\right)},~\text{for $K\geq1$}.~~~ \label{ConK1}
\end{subnumcases}

We then derive the second term $\Pr\left(X_0\in{S_1}\big|k=K,k\geq{1}\right)$ in \eqref{qqjj}.
As $k\sim{\text{Poi}(\mu)}$ and the two sectors are equally partitioned, we have $k_1\sim\text{Poi}({\mu}/{2})$ and $k_2\sim\text{Poi}({\mu}/{2})$. Using the property of Poisson distribution, $k_1$ conditioned on $k_1+k_2=k$ follows a Binomial distribution. That is, $k_1\sim\text{Binom}(k,\frac{1}{2})$, which indicates that each of the $k$ i.i.d. MUs falls into the two sectors with equal probabilities.
As the typical users is chosen from these MUs, we thus have
\begin{align}
\Pr\left(X_0\in{S_1}\big|k=K,k\geq{1}\right)=\frac{1}{2}.
\label{halff}
\end{align}

The third term $\Pr\left(U=U_j\big|X_0\in{S_1},k=K,k\geq{1}\right)$ in \eqref{qqjj} can be deduced as follows. Given $X_0\in{S_1}$ and $k_1+k_2=K$, the UAV is displaced to $(U_1,h)$  if at most $\lfloor{\frac{K-1}{2}\rfloor}$ out of the rest of $K-1$ users fall into the sector $S_2$, where the floor function ensures that $k_2$ is an integer. We thus have \eqref{qj1}.
Conditioned on $X_0\in{S_1}$ and $k_1+k_2=K$, the UAV is displaced to $(U_2,h)$ if at most $\lfloor{\frac{K-1}{2}\rfloor}-1$  out of the rest of $K-1$ users are inside $S_1$. We thus have \eqref{qj2}. The UAV chooses the center displacement position of $(U_0,h)$ if neither of the above cases happens.
Since \begin{align}
\sum_{j=0}^{2}q_j=\sum_{j=0}^{2}\Pr\left(X_0\in{S_1},U=U_j\big|k\geq{1}\right)=\Pr\left(X_0\in{S_1}\big|k\geq{1}\right)=\frac{1}{2},\label{halfff}
\end{align}
 we have \eqref{qj3}.
\end{IEEEproof}

Based on \eqref{qj1} and \eqref{qj2}, we can easily deduce that $q_1>q_2$.
To obtain more insights on how the average traffic load affects the joint probability $q_j$, we examine some asymptotic properties in the following corollary.
\begin{corollary}
As the average traffic load in the cell goes to zero (i.e., $\mu\rightarrow{0}$), the displacement probabilities of the UAV satisfy $q_1\rightarrow{\frac{1}{2}}$, $q_2\rightarrow{0}$ and $q_0\rightarrow{0}$.
As $\mu\rightarrow{\infty}$, the displacement probabilities satisfy $q_1\rightarrow{\frac{1}{4}}$, $q_2\rightarrow{\frac{1}{4}}$ and $q_0\rightarrow{0}$.
\end{corollary}
\begin{IEEEproof}
By further derivations, we can simplify the third term in \eqref{qqjj} for $j=1$ as
\begin{subnumcases}
{\Pr\left(U=U_1\big|X_0\in{S_1},k=K,k\geq{1}\right)=\nonumber}
\frac{1}{2}+\frac{C_{K-1}^{\frac{K-1}{2}}}{2^{K}},~\text{if}~K~\text{is odd}\label{odd1}\\
\frac{1}{2},~\text{if}~K~\text{is even}\label{even1}
\end{subnumcases}
and for $j=2$ as
\begin{subnumcases}{\Pr\left(U=U_2\big|X_0\in{S_1},k=K,k\geq{1}\right)=\nonumber}
\frac{1}{2}-\frac{C_{K-1}^{\frac{K-1}{2}}}{2^{K}},~\text{if}~K~\text{is odd} \label{odd2}\\
\frac{1}{2}-\frac{C_K^{K/2}}{2^{K}},~\text{if}~K~\text{is even}.\label{even2}
\end{subnumcases}

As $\mu\rightarrow{0}$, we have $e^{-\mu}=1-\mu$. According to \eqref{ConK1},
we deduce that $\lim\limits_{\mu\rightarrow{0}}\Pr\left(k=1\big|k\geq{1}\right)={1}$.
For $k=1$, we can deduce from \eqref{odd1} and \eqref{odd2} that
$\lim\limits_{\mu\rightarrow{0}}\Pr\left(U=U_1\big|X_0\in{S_1},k=1\right)=\frac{1}{2}+\frac{C_{0}^{0}}{2^1}=1$ and $\lim\limits_{\mu\rightarrow{0}}\Pr\left(U=U_2\big|X_0\in{S_1},k=1\right)=\frac{1}{2}-\frac{C_{0}^{0}}{2^1}=0.$
By substituting them and \eqref{halff} into \eqref{qqjj}, we have $\lim\limits_{\mu\rightarrow{0}}q_1=\frac{1}{2}$.
and $\lim\limits_{\mu\rightarrow{0}}q_2={0}$. And we thus have $\lim\limits_{\mu\rightarrow{0}}q_0={0}$ based on \eqref{qj3}.

As $\mu\rightarrow\infty$, we are almost sure that
$\lim\limits_{\mu\rightarrow{\infty}}\Pr\left(k\rightarrow\infty\big|k\geq{1}\right)=1$.
For $k\rightarrow{\infty}$, we can deduce from \eqref{odd1} and \eqref{even1} that
$\lim\limits_{\mu\rightarrow{\infty}}\Pr\left(U=U_1\big|X_0\in{S_1},k\rightarrow\infty\right)=\frac{1}{2}$
and can obtain from \eqref{odd2} and \eqref{even2} that
$\lim\limits_{\mu\rightarrow{\infty}}\Pr\left(U=U_2\big|X_0\in{S_1},k\rightarrow\infty\right)=\frac{1}{2}-\frac{1}{2^{\infty}}$.
By substituting them and \eqref{halff}
into \eqref{qqjj}, we have
$\lim\limits_{\mu\rightarrow{\infty}}q_1
=\frac{1}{4}$ and  $\lim\limits_{\mu\rightarrow{\infty}}q_2
={\frac{1}{4}}$.
Thus we have $\lim\limits_{\mu\rightarrow{\infty}}q_0={0}$ based on \eqref{qj3}.
\end{IEEEproof}
\begin{remark}
Based on Corollary 1, the typical MU$_0$'s location affects the UAV's displacement location significantly when the average traffic load is low. As $\mu\rightarrow{0}$,
MU$_0$ is very likely to be the only MU in the cell given $k\geq{1}$. Conditioned on $X_0\in{S_1}$, the UAV will surely move to $U=U_1$.
Since the probability of $X_0\in{S_1}$ is $1/2$, the joint probability of the events of $X_0\in{S_1}$ and $U=U_1$ is thus $1/2$.
For very high traffic load, the impact of MU$_0$'s location on the UAV's displacement location is trivial. As $\mu\rightarrow{\infty}$, the events of $X_0\in{S_1}$ and $U=U_1$ are almost independent and each happens with the probability of $1/2$. The joint probability of the two events is thus $1/4$.
\end{remark}

\subsection{Average Throughput of MU$_0$}
We now derive the conditional average throughput $\omega_j$.
Conditioned on that the UAV moves to $U_j$ and MU$_0$ is within $S_1$, the average throughput is given by
\begin{align}
\omega_{j}
=\int_{-R}^{0}\log\left[1+\frac{a}{\left(|X_0-U_j|^2+h^2\right)}\right]f(X_0|X_0\in{S_1})dX_0,\label{o1}
\end{align}
where $f(X_0|X_0\in{S_1})=\frac{1}{R}$  since $X_0$ is uniformly distributed in the interval of $[-R,0]$.

Substituting all $q_j$ in \eqref{qj1}, \eqref{qj2} and \eqref{qj3} and $\omega_j$ in \eqref{o1}~for $j=0,1,2$
into \eqref{ECCC}, the average throughput of MU$_0$ is given by the following proposition.
\begin{proposition}
In the UAV-aided 1D  mobile network, the average throughput of the typical MU$_0$ under the adaptive UAV deployment scheme in (1) is given by
\begin{align}
\mathbb{E}\left[C\big|k\geq{1}\right]=2\left[q_{1}\zeta+(q_{1}-q_{2})\kappa+q_{2}\xi+q_{0}\vartheta\right],\label{EEC}
\end{align}
where
$\zeta
=\frac{2h}{R}\arctan\left(\frac{(\beta-1)R}{h}\right)-\frac{2\sqrt{a+h^2}}{R}\arctan\left(\frac{(\beta-1)R}{\sqrt{a+h^2}}\right)
-(\beta-1)\log\left(1+\frac{a}{h^2+(\beta-1)^2R^2}\right)$,
$\kappa
=-\frac{2h}{R}\arctan\left(\frac{\beta{R}}{h}\right)+\frac{2\sqrt{a+h^2}}{R}\arctan\left(\frac{\beta{R}}{\sqrt{a+h^2}}\right)
+\beta\log\left(1+\frac{a}{h^2+\beta^2R^2}\right)$,
$\xi
=-\frac{2h}{R}\arctan\left(\frac{(\beta+1)R}{h}\right)+\frac{2\sqrt{a+h^2}}{R}\arctan\left(\frac{(\beta+1)R}{\sqrt{a+h^2}}\right)
+(\beta+1)\log\left(1+\frac{a}{h^2+(\beta+1)^2R^2}\right)$
and
$\vartheta
=\frac{1}{R}[-2h\arctan\left(\frac{R}{h}\right)+2\sqrt{a+h^2}$
$\arctan\left(\frac{R}{\sqrt{a+h^2}}\right)]
+\log\left(1+\frac{a}{h^2+R^2}\right)$.
\end{proposition}

In the following corollary, we prove that $\mathbb{E}\left[C\big|k\geq{1}\right]$ in Proposition~2 above is concave in $\beta$.
\begin{corollary}
$\mathbb{E}\left[C\big|k\geq{1}\right]$ is strictly concave in the displacement factor $\beta$ for $\beta\in[0,1]$.
\end{corollary}
\begin{IEEEproof}
The average throughput in \eqref{EEC} is a linear combination of $\zeta$, $\kappa$, $\xi$, $\vartheta$ and the probabilities $q_j$ ($j=0,1,2$). Since $\vartheta$ and $q_j$ are not functions of $\beta$, we just check the second derivatives of $\zeta$, $\kappa$ and $\xi$ with respect to $\beta$, i.e.,
$\frac{\partial^2\zeta}{\partial\beta^2}=-\frac{2a(1-\beta)R^2}{\left(h^2+(1-\beta)^2R^2\right)\left(a+h^2+(1-\beta)^2R^2\right)}$,
$\frac{\partial^2\kappa}{\partial\beta^2}=-\frac{2a\beta{R}^2}{\left(h^2+\beta^2R^2\right)\left(a+h^2+\beta^2R^2\right)}$,
and
$\frac{\partial^2\xi}{\partial\beta^2}
=-\frac{2a(1+\beta)R^2}{\left(h^2+(1+\beta)^2R^2\right)\left(a+h^2+(1+\beta)^2R^2\right)}$.
Since $0\leq\beta\leq1$, we have $\frac{\partial^2\zeta}{\partial\beta^2}\leq{0}$, $\frac{\partial^2\kappa}{\partial\beta^2}\leq{0}$ and $\frac{\partial^2\xi}{\partial\beta^2}<0$, where the equality signs of the first and second terms hold if $\beta=1$ and $\beta=0$, respectively.
Based on Proposition 1, we have $q_1>0$, $q_2\geq0$ and $q_1>{q}_2$.
We thus have
$\frac{\partial^2\mathbb{E}[C]}{\partial\beta^2}
=2\left[q_1\frac{\partial^2\zeta}{\partial\beta^2}
+(q_1-q_2)\frac{\partial^2\kappa}{\partial\beta^2}+q_2\frac{\partial^2\xi}{\partial\beta^2}\right]
<0$
for all $\beta\in[0,1]$. Therefore, $\mathbb{E}[C]$ is strictly concave in $\beta$.
\end{IEEEproof}

\begin{remark}
Intuitively, Corollary 2 tells a fundamental tradeoff in the displacement distance design. If the displacement distance is small,  the UAV cannot efficiently serve the users in the target sector with more MUs. If the displacement distance is large, the MUs in the other sector with less MUs will suffer from great throughput degradation.
Thus, designing the optimal displacement distance is of critical importance to maximize the network average throughput.
\end{remark}

\subsection{Optimal Displacement Factor}
According to Corollary 2,  there is a unique optimal displacement factor $\beta$ that maximizes the average throughput.
The optimization problem is
\begin{eqnarray}
   &&\text{P1}:\max\limits_{0\leq{\beta}\leq{1}}~~{\mathbb{E}\left[C\big|k\geq{1}\right]}.
\end{eqnarray}
By solving P1, we have the following proposition.\footnotemark
\footnotetext{Note that the final formula of $\beta^*$ as the root of \eqref{prop3} in Proposition 3  cannot be obtained explicitly but can be solved numerically via root-finding algorithms, e.g., bisection search. Similar approaches will be applied to find the solutions of $\beta^*$ to other implicit equations in Propositions 7 and 8 in Section V.}
\begin{proposition}
The optimal displacement factor $\beta^*$ that maximizes the average throughput $\mathbb{E}\left[C\big|k\geq{1}\right]$ is the unique solution to
\begin{align}
q_1\varrho_1
-q_2\varrho_2=0.\label{prop3}
\end{align}
where $\varrho_1=\log\left(\frac{1+\frac{a}{h^2+\beta^2R^2}}{1+\frac{a}{h^2+(\beta-1)^2R^2}}\right)$
and $\varrho_2=\log\left(\frac{1+\frac{a}{h^2+\beta^2R^2}}{1+\frac{a}{h^2+(\beta+1)^2R^2}}\right)$.
The optimal $\beta^*$ decreases with average traffic load $\mu$ in the cell. Furthermore, we have $\beta^*\rightarrow\frac{1}{2}$ as $\mu\rightarrow{0}$ and
$\beta^*\rightarrow0$ as $\mu\rightarrow{\infty}$, respectively.
\end{proposition}
\begin{IEEEproof}
Due to the concavity of the objective function of P1 as shown in Corollary 2, it is enough to check  the first-order condition.
We thus have \eqref{prop3} or equivalently  $q_1/q_2=\varrho_2/\varrho_1$.
We can prove that $\varrho_2/\varrho_1$ is positive for $\beta^*\in[0,0.5]$ and negative for $\beta^*\in[0.5,1]$, respectively. Since $q_1/q_2>0$, the feasible $\beta^*$ should be within the regime of $[0,0.5]$. In this regime, $\varrho_2/\varrho_1$ is monotonically increasing in $\beta^*$.
According to \eqref{qj1} and \eqref{qj2},  $q_1$ decreases with $\mu$, and $q_2$ increases with $\mu$, respectively. Thus, $q_1/q_2$ decreases with $\mu$.
As a result, $\beta^*$ decreases with $\mu$.

We then prove the asymptotic results of $\beta^*$ in terms of $\mu$. As $\mu\rightarrow{0}$, we have $q_1\rightarrow{1/2}$ and $q_2\rightarrow{0}$ based on Corollary 1. Substituting them into \eqref{prop3}, we have $\beta^2=(\beta-1)^2$ and thus have $\beta^*\rightarrow{1/2}$. As $\mu\rightarrow{\infty}$, we have $q_1\rightarrow{1/4}$ and $q_2\rightarrow{1/4}$ based on Corollary 1. Substituting them into \eqref{prop3}, we have $(\beta-1)^2=(\beta+1)^2$ and thus have $\beta^*\rightarrow{0}$.
\end{IEEEproof}
\begin{remark}
As $\mu\rightarrow{0}$, MU$_0$ (if any) is very likely to be the only user in the network. To best serve this user in average sense without knowing its exact location, it is best for the UAV to move to the center of $S_j$ and the corresponding $\beta^*=0.5$. As $\mu\rightarrow\infty$, users are balanced in different sectors as in the average sense and the UAV should stay in the center with  $\beta^*=0$.
\end{remark}

We further compare the maximum average throughput $\mathbb{E}[C(\beta^*)|k\geq{1}]$ with the average throughput $\mathbb{E}[C_0|k\geq{1}]$ in the non-adaptive scheme.
For the non-adaptive scheme, the UAV is located at the origin, which is a special case of the proposed scheme by using $\beta=0$.
Since $\mathbb{E}[C|k\geq{1}]$ is concave in $\beta$, we obtain the following corollary.
\begin{corollary}
The maximum average throughput in the proposed adaptive UAV deployment scheme  with the optimal $\beta^*$  outperforms that of the non-adaptive scheme, i.e., $\mathbb{E}[C(\beta^*)|k\geq{1}]\geq\mathbb{E}[C_0|k\geq{1}]$.
\end{corollary}

\section{Successful Transmission Probability Maximization with 1D Adaptive UAV Deployment}

In this section, we study the optimal UAV deployment for a delay-limited constant-rate transmission application (voice call or on-line gaming), where the UAV/MUs transmit with a constant rate in the downlink/uplink  and the transmission is successful if the instantaneous SNR at the receiver is greater than the target threshold of $\gamma_{th}$.
In the following, we design the optimal $\beta$
to maximize the probability that the UAV successfully transmits/receives message to/from the typical MU$_0$ under the proposed 1D adaptive UAV deployment scheme in (1). Intuitively, the design of $\beta$ is related to the SNR threshold. For example, when the SNR target is low, the UAV has relatively large coverage and does not need to move far away from the cell center.
To obtain the insight on how the SNR threshold affects the UAV deployment distance from an average perspective, we adopt the same $\beta$ across different MU number/location  realizations in the following discussions. At the end of this section, we will show that this is exactly equivalent to updating with a different $\beta$ in each realization due to the unique successful  transmission (success in short) probability metric.

\subsection{Tractable Analysis of Success Probability}
Conditioned on $k\geq{1}$, the success probability of MU$_0$ is defined as the probability that the received SNR $\gamma$ in \eqref{gamma} is no smaller than the target SNR threshold $\gamma_{th}$, i.e.,
\begin{align}
p=\Pr\left[\gamma\geq\gamma_{th}\big|k\geq{1}\right]
=\Pr\left[|X_0-{U}|\leq{\rho}\big|k\geq{1}\right],\label{psuc}
\end{align}
where we denote
$\rho=\sqrt{\frac{a}{\gamma_{th}}-h^2}$ with $a=P_t\theta/\sigma^2$
as the UAV's coverage radius that describes the maximum horizontal distance between MU$_0$ and UAV for achieving the target SNR performance. Given the UAV is in the position of $(U_j,h)$ in Fig.~1, the corresponding coverage region is thus $[U_j-\rho,U_j+\rho]$.
Since we always have $p=1$ once $\rho\geq{R}$ even in the non-adaptive scheme, in the sequel, we focus on the scenario of $0<\rho<{R}$.

Similar to the derivation of the average throughput in \eqref{ECCC}, we have
\begin{align}
p=2\sum\limits_{j=0}^{2}q_j\eta_{j},
\label{totalpp}
\end{align}
where the displacement probability $q_j$ is given in Proposition 1, and we further denote
$\eta_j=\Pr\left[|X_0-U|\leq{\rho}\big|U=U_j,X_0\in{S_1}\right]$
as the conditional success probability of MU$_0$.
By recalling the property of $\sum_{j=0}^{2}q_j=\Pr\left(X_0\in{S_1}\big|k\geq{1}\right)=\frac{1}{2}$ given in \eqref{halfff}, the overall success probability of MU$_0$ in \eqref{totalpp} is given by the following proposition.
\begin{proposition}
Under the adaptive 1D scheme in (1), the success probability of the typical MU$_0$ is
\begin{subnumcases}{p=\nonumber}
\frac{\rho}{R}+2\beta\left(q_1-q_2\right),~~~~~~~~~~~~\text{if}~{\beta}\leq\frac{\rho}{R}~~\label{P41}\\
\frac{4\rho}{R}q_1+\frac{2\rho}{R}q_0,~~~~~~~~~~~~~~~~~~\text{if}~\frac{\rho}{R}\leq{\beta}\leq1-{\frac{\rho}{R}}~~\label{P42}\\
2\left[\left(1-\beta+\frac{\rho}{R}\right)q_1+\frac{\rho}{R}q_0\right],~\text{if}~1-\frac{\rho}{R}\leq{\beta}\leq{1}~~\label{P43}
\end{subnumcases}
for $0<\rho\leq\frac{R}{2}$ and
\begin{subnumcases}{p=\nonumber}
\frac{\rho}{R}+2\beta\left(q_1-q_2\right),~~~~~~~~~~~~\text{if}~{\beta}\leq1-\frac{\rho}{R}\label{P44}\\
\frac{\rho}{R}+2\left[(1-\frac{\rho}{R})q_1-\beta{q}_2\right],~~\text{if}~1-\frac{\rho}{R}\leq{\beta}\leq{\frac{\rho}{R}}\label{P45}\\
2\left[\left(1-\beta+\frac{\rho}{R}\right)q_1+\frac{\rho}{R}q_0\right],~\text{if}~\frac{\rho}{R}\leq{\beta}\leq{1}\label{P46}
\end{subnumcases}
for $\frac{R}{2}<\rho\leq{1}$.
\end{proposition}

\subsection{Optimal Displacement Factor}
In this subsection, we design the optimal displacement factor $\beta$ to maximize the success probability of the typical MU$_0$ given in Proposition 4. That is,
\begin{eqnarray}
   &&\text{P2}:\max\limits_{0\leq{\beta}\leq{1},~\rho+\beta{R}\leq{R}}~~{p(\beta)}
\end{eqnarray}
Note, the constraint of $\rho+\beta{R}\leq{R}$ ensures the UAV does not move its coverage region outside the cell.
Since $q_1>q_2$, we can easily prove that the success probability $p(\beta)$ in Proposition~4 is a concave function of $\beta$. By solving P2, we have the following proposition.
\begin{proposition}
If $0<\rho\leq\frac{R}{2}$, any displacement factor
$\beta^*\in\left[\frac{\rho}{R},1-\frac{\rho}{R}\right]$ is optimal.
If $\frac{R}{2}<\rho\leq{R}$, the unique optimal displacement factor is
$\beta^*=1-\frac{\rho}{R}$.
\end{proposition}
We can prove that the optimal $\beta^*$ in Proposition 5 holds even if we allow the UAV to maximize the success probability in each realization according to the exact values of $k_1=K_1$ and $k_2=K_2$ (except for the symmetric case of $K_1=K_2$ where we always have $\beta^*=0$). For example, if $K_1>K_2$, the success probability $p$ can be obtained by substituting $q_0=0$, $q_1=K_1/(K_1+K_2)$ and $q_2=K_2/(K_1+K_2)$ into Proposition 4 (the derivation is similar to that in Appendix A). By solving P2, $\beta^*$ is given by Proposition 5 as well. The similar proof applies for $K_1<K_2$.

We now assume $K_1>K_2$ for the UAV's displacement to sector $S_1$ and present the intuitive explanations for Proposition 5.
\begin{itemize}
\item If $0<\rho\leq{\frac{R}{2}}$, the coverage region of the UAV is not large enough to cover the whole sector of $S_1$. In this case, without knowing the specific locations of the users, the optimal strategy for the UAV is to
    keep its coverage region $[U_1-\rho, U_1+\rho]$ within $S_1$.
    Thus, any UAV location point that satisfies the above condition is optimal.
\item  If $\frac{R}{2}<\rho\leq{R}$,
the UAV
      is able to not only provide full coverage for the users in $S_1$ but also cover $S_2$ as much as possible.
      Note that the UAV still has more preference for serving the users in $S_1$ than $S_2$ given most of the users are in $S_1$. To avoid moving its coverage region outside the cell and save coverage for $S_1$, the left-most coverage should just reach the left boundary of $S_1$.
\end{itemize}

We further compare the maximum success probability $p(\beta^*)$ with the success probability $p_0$ in the non-adaptive scheme. Based on Propositions 4 and 5, $p(\beta)$ is increasing in $\beta\in[0,\beta^*]$ for both cases of $\rho\in[0,R/2]$ and $\rho\in[R/2,R)$, we thus have the following corollary.
\begin{corollary}
The maximum success probability with the proposed adaptive scheme in (1) strictly outperforms that of the non-adaptive scheme, i.e., $p(\beta^*)>{p}_0$, for any $\rho\in[0,R)$.
\end{corollary}

\section{Extension of Adaptive UAV Deployment for 2D User Network}

In this section, we extend the design and analysis to the 2D MU network to maximize the average throughput and success probability of the typical user, respectively.

\begin{figure}[t!]
    \begin{center}
        \includegraphics[width=0.4\columnwidth]{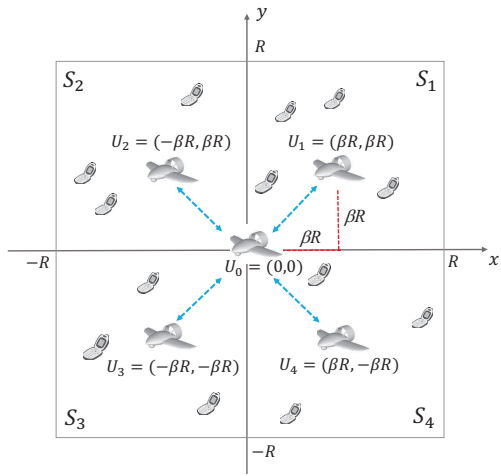}
        \caption{A UAV-aided two-dimensional network. The UAV is adaptively dispatched to one of the five candidate locations of $(U_j,h)$, where $U_0=(0,0)$, $U_1=(\beta{R},\beta{R})$, $U_2=(-\beta{R},\beta{R})$, $U_3=(-\beta{R},-\beta{R})$, and $U_4=(\beta{R},-\beta{R})$.}
    \end{center}
\end{figure}

We plot the UAV-aided 2D network in Fig. 2. The MUs follow a homogeneous PPP $\{W_m\}$ with the spatial density of $\lambda$, where $W_m$ is the coordinate of MU$_m$ in $\mathcal{R}^2$ on the ground. The UAV aims at serving the users that are inside its target cell which is modeled as a square region
with the width of $2R$.
The initial location (or that of the non-adaptive scheme) of the UAV is  at the center of the square with the height of $h$, i.e., $(U_0,h)=(0,0,h)$. We equally divide the target cell into four sectors $S_i$ ($i=1,2,3,4$). In one realization, we denote the number of MUs in sector $S_i$ as $k_i$ and the total number of users in the cell as $k$. Similar to the 1D case, we consider that the UAV knows and adapts to only the number of MUs in each sector without knowing the specific locations of the MUs.

Different from the 1D network in Fig. 1, the UAV deployment in the 2D network involves not only the displacement distance but also multiple moving directions.
In this section, we
adopt the same majority-vote rule as in the 1D case for simplicity and assume that the UAV has limited moving direction choices, i.e., it moves only along the two diagonals of the square cell.
Hence, the UAV adapts its location $U$
by choosing one of the five displacement positions $(U_j,h)=(U_{j,x},U_{j,y},h)$ (for $j=0,1,2,3,4$) as its new displacement location $(U,h)$, i.e.,
\begin{subnumcases}
{U=}
U_j,~\text{if $k_j>\max\limits_{i\neq{j}}(k_i)$} \label{YY1}\\
U_0,~\text{otherwise}, \label{YY0}
\end{subnumcases}
where $U_0=(0,0)$, $U_1=\left(\beta{R},\beta{R}\right)$, $U_2=\left(-\beta{R},\beta{R}\right)$, $U_3=\left(-\beta{R},-\beta{R}\right)$ and $U_4=\left(\beta{R},-\beta{R}\right)$ as shown in Fig. 2.
To obtain the insights on how the traffic load and target SNR threshold affect the deployment distance, we adopt the same $\beta$ across different realizations as in the 1D network case.
We will extend and also compare it with other  schemes in Section VI.
In the following, we will
derive the optimal $\beta^*$ to maximize the objectives of the average throughput and success probability, respectively.

\subsection{2D Adaptive UAV Deployment for Average Throughput Maximization}
In this subsection, we analyze the optimal displacement factor for average throughput maximization under the proposed 2D adaptive UAV deployment in (25).

To evaluate the performance of an arbitrary MU, we randomly select a typical MU from the existing MUs (if any) inside the cell, denoted by MU$_0$, where its 2D ground location is denoted by $W_0=(X_0,Y_0)$ in $\mathcal{R}^2$.
The throughput of MU$_0$ is given by
\begin{align}
C=\log\left(1+\gamma\right)
=\log\left(1+\frac{a}{\|W_0-U\|^2+h^2}\right),\label{Cnew}
\end{align}
where both
$W_0$ and $U$ are 2D random variables that are correlated with each other.

\subsubsection{Tractable Analysis of Average Throughput}
To derive the long-term average throughput of MU$_0$, we average over the five location candidates of the UAV, i.e., $U_j$ ($j=0,1,2,3,4$) and the four location regions of the MU$_0$, i.e., $W_0\in{S_i}$ ($i=1,2,3,4$). According to the probability chain rule, the average throughput in \eqref{Cnew} assuming at least one MU's presence is given by
\begin{align}
\mathbb{E}[C\big|k\geq{1}]
&=\sum\limits_{j=0}^{4}\sum\limits_{i=1}^{4}\mathbb{E}\left[C\big|U=U_j,W_0\in{S_i},k\geq{1}\right]
\Pr\left(U=U_j,W_0\in{S_i}\big|k\geq{1}\right).\label{ECCC1}
\end{align}
Since the analysis is symmetric for all the four cases of $W_0\in{S_i}$, we can replace $\sum_{i=1}^{4}$ in \eqref{ECCC1} by $4$ and just focus on the analysis of case $W_0\in{S_1}$. As such, we rewrite \eqref{ECCC1} as
\begin{align}
\mathbb{E}[C\big|k\geq{1}]&=4\sum\limits_{j=0}^{4}q_j\mathbb{E}\left[C\big|U=U_j,W_0\in{S_1},k\geq{1}\right],~~\label{ECCC2}
\end{align}
where we denote the joint probability that the UAV is displaced $U_j$ and MU$_0$ is inside $S_1$ as
\begin{align}
q_j&=\Pr\left(U=U_j,W_0\in{S_1}\big|k\geq{1}\right)=\Pr\left(k_j>\max\limits_{i\neq{j}}(k_i),W_0\in{S_1}\big|k\geq{1}\right).
\label{qjj}
\end{align}

The derivation of
\eqref{qjj} is challenging due to the correlation between $U=U_j$ and $W_0\in{S_i}$. This is because that MU$_0$ is one of the users whose locations affect the displacement decision of the UAV.
Moreover, the events of $k_j>k_i$ and $k_j>k_n$ (for any other sector $n\neq{i}$) are also correlated,
so that we cannot decompose the events of $k_j>\max_{i\neq{j}}(k_i)$ in \eqref{qjj} into $M-1$ independent events of $k_j>k_i$.
We thus adopt the multi-layer convolution to solve this problem.
To generalize the results, we
consider the general $M$-sector case
and will derive the average throughput using $M=4$ (as in Fig. 2) for tractable results later.
We obtain $q_j$ in the following.
\begin{proposition}
Conditioned on $k\geq{1}$ in the $M$-sector MU network, the joint probability that the UAV chooses the displacement position of $U=U_j$ (for $j=0,1,2,3,4$) and the typical MU$_0$ is inside the sector $S_1$ in an $M$-sector MU network is given by
\begin{subnumcases}{q_j=\nonumber}
\frac{e^{-\mu}}{1-e^{-\mu}}
\sum_{t_M=1}^{\infty}\cdots\sum_{t_2=1}^{\infty}\sum_{t_1=\max(t_2,\cdots,t_M)}^{\infty}
\frac{t_1\left(\frac{\mu}{M}\right)^{Mt_1-\sum_{i=2}^{M}t_i}}{(Mt_1-\sum_{i=2}^{M}t_i)t_1!\prod_{i=2}^{M}(t_1-t_i)!}
,~\text{for}~j=1~~\nonumber\\
\frac{e^{-\mu}}{1-e^{-\mu}}\sum_{t_M=1}^{\infty}\cdots\sum_{t_2=1}^{\infty}
\sum_{t_1=\max(t_2+1,t_3,\cdots,t_M)}^{\infty}
\frac{(t_1-t_2)\left(\frac{\mu}{M}\right)^{Mt_1-\sum_{j=2}^{M}t_i}}{(Mt_1-\sum_{i=2}^{M}t_i)t_1!\prod_{i=2}^{M}(t_1-t_i)!},~\text{for}~j\neq{0,1}\nonumber\\
\frac{1}{M}-\sum\limits_{i=1}^{M}q_i,~\text{for}~j=0,\nonumber
\end{subnumcases}
where $\mu=4R^2\lambda$ is the average number of users in the cell.
As the average user number in the cell $\mu\rightarrow{0}$, we have $q_1\rightarrow{1/M}$ and $q_j\rightarrow{0}$ for any $j\neq{1}$.
As $\mu\rightarrow{\infty}$, we have $q_0\rightarrow{0}$ and $q_j\rightarrow{{1}/{M^2}}$ for any $j\neq{0}$.
\end{proposition}
\begin{IEEEproof}
See Appendix B.
\end{IEEEproof}
The insight of Proposition 6 is similar to that of Remark 1 by replacing $1/2$ in the 1D's two sector case by $1/M$ here in 2D, though we have adopted two different methods to derive the joint probabilities for the 1D and 2D networks.

Conditioned on the joint event that the UAV is displaced to $U_j$ and MU$_0$ is within $S_1$, the average throughput 
of the typical user for the case of $M=4$ is given by
\begin{align}
\mathbb{E}[C\big|U=U_j,W_0\in{S_1}]
=\int_{0}^{R}\int_{0}^{R}\log\left[1+\gamma\big|U=U_j,W_0\in{S_1}\right]
\frac{1}{R^2}dX_0dY_0.
&\label{omega}
\end{align}
By substituting $q_j$ in Proposition 6 and \eqref{omega}
for $j=0,1,2,3,4$  into \eqref{ECCC2}, we can obtain the expression of $\mathbb{E}[C\big|k\geq{1}]$, for which the exact expression is omitted for brevity.

\subsubsection{Optimal Displacement Factor for Average Throughput Maximization}
Similar to P1 for the 1D network, we now derive the optimal displacement distance for average throughput maximization for the 2D network.
We can prove that $\mathbb{E}[C\big|k\geq{1}]$ is concave in $\beta$ and there is a unique optimal displacement factor $\beta$ by solving the first-order condition
${d{\mathbb{E}\left[C\big|k\geq{1}\right]}}/{d{\beta}}
=0$.
To further simplify the above equation,
we define three special functions of
$f(z,v)=\sqrt{h^2+z^2R^2}\arctan\left(\frac{vR}{\sqrt{h^2+z^2R^2}}\right)$
$g(z,v)=\sqrt{a+h^2+z^2R^2}\arctan\left(\frac{vR}{\sqrt{a+h^2+z^2R^2}}\right)$
and
$s(z,v)=zR\log\left(1+\frac{a}{h^2+z^2R^2+v^2R^2}\right)$.
Due to the symmetric properties, we use $q_2$ to represent all identical $q_j$ for $j\neq{0,1}$ without loss of generality. The optimal $\beta^*$ is given in the following proposition.
\begin{proposition}
Under the proposed 2D adaptive UAV deployment scheme in (25), the UAV's optimal displacement factor $\beta^*$ that maximizes the average throughput of the typical MU$_0$ is the unique solution to
\begin{align}
&(q_1-q_2)[2f(1-\beta,\beta)-2f(\beta,\beta)-2f(\beta,1-\beta)
+2g(\beta,\beta)+2g(\beta,1-\beta)-2g(1-\beta,\beta)\nonumber\\
&+s(\beta,\beta)+s(1-\beta,\beta)-s(\beta,1-\beta)]
+2q_1[f(1-\beta,1-\beta)-g(1-\beta,1-\beta)\nonumber\\
&-s(1-\beta,1-\beta)]+q_2[2f(1-\beta,1+\beta)-2f(1+\beta,1-\beta)-2f(1+\beta,1+\beta)\nonumber\\
&+2g(1+\beta,1-\beta)+2g(1+\beta,1+\beta)
-2g(1-\beta,1+\beta)+s(1+\beta,1+\beta)\nonumber\\
&+s(1-\beta,1+\beta)-s(1+\beta,1-\beta)]=0.
\label{propp}
\end{align}
As $\mu\rightarrow{0}$, we have $\beta^*\rightarrow{0.5}$; and as $\mu\rightarrow{\infty}$, we have $\beta^*\rightarrow{0}$.
\end{proposition}
\begin{IEEEproof}
We can simplify ${d{\mathbb{E}\left[C\big|k\geq{1}\right]}}/{d{\beta}}=0$ to obtain \eqref{propp}.
As $\mu\rightarrow{0}$, we have $q_1\rightarrow{1/4}$ and $q_i\rightarrow0$ $\forall$ $i\neq{1}$ according to Proposition 6 for $M=4$. In this case, 
\eqref{propp} holds if and  only if $\beta=1-\beta$, and we thus have $\beta^*={0.5}$.
As $\mu\rightarrow\infty$, we have $q_j\rightarrow\frac{1}{16}$ $\forall$ $j\neq{0}$ according to Proposition 6 for $M=4$. In this case, \eqref{propp}
holds if and only if $1-\beta=1+\beta$. We thus have $\beta^*={0}$.
\end{IEEEproof}

We see that the above asymptotic results of the optimal $\beta^*$ is similar to Proposition 3 of the 1D network, where the UAV moves to the center of the chosen sector when the user density is small and the UAV displacement
becomes mild when the user density is large.

\subsection{2D Adaptive UAV Deployment for Success Probability Maximization}
In this subsection, we derive the optimal displacement distance for success probability maximization in 2D and discuss how it changes with the target SNR threshold and average traffic load.
Similar to \eqref{ECCC2}, the success probability of MU$_0$ in the four-sector 2D network is
\begin{align}
&p
=4\sum\limits_{j=0}^{4}q_j\eta_{j},\label{totalppp}
\end{align}
where $q_j$ is given in Proposition 6
and
$\eta_j=\Pr\left[\|W_0-U\|\leq{\rho}\big|U=U_j,W_0\in{S_1}\right]$.
For the 2D MU network, the UAV's coverage region is no longer a line interval but  a circular disk that is centered at $U$ with radius $\rho=\sqrt{\frac{a}{\gamma_{th}}-h^2}$.
We can prove that $p$ is concave in $\beta$ and derive the optimal $\beta^*$  similar to P2.
We replace $\rho+\beta{R}<R$ by $\rho+\sqrt{2}\beta{R}<\sqrt{2}R$ for the 2D network to ensure that the UAV does not waste its coverage outside the cell. We use $q_2$ to replace any identical $q_j$ ($\forall~j\neq{0,1}$) without loss of generality.
\begin{proposition}
The UAV's  optimal displacement factor $\beta^*$ that maximizes the success probability for the typical MU$_0$ in the 2D MU network is given by
\begin{subnumcases}{\beta^*=\nonumber}
\beta_0^*,~\text{if}~\rho\in\left[0,\frac{R}{2}\right)\\
\beta_1^*,~\text{if}~\rho\in\left[\frac{R}{2},\frac{\sqrt{2}Rq_1(\sqrt{q_1^2+2q_1q_2-q_2^2}-q_1)}{q_2(2q_1-q_2)}\right)\label{bb1}\\
\beta_2^*,~\text{if}~\rho\in\left[\frac{\sqrt{2}Rq_1(\sqrt{q_1^2+2q_1q_2-q_2^2}-q_1)}{q_2(2q_1-q_2)},\frac{R\sqrt{q_1^2+q_2^2}}{\sqrt{2}q_1}\right)\label{bb2}\\
\beta_3^*,~\text{if}~\rho\in\left[\frac{R\sqrt{q_1^2+q_2^2}}{\sqrt{2}q_1},\frac{2R}{1+\frac{q_1+q_2}{\sqrt{2(q_1^2+q_2^2)}}}\right)\label{bb3}\\
\beta_4^*,~\text{if}~\rho\in\left[\frac{2R}{1+\frac{q_1+q_2}{\sqrt{2(q_1^2+q_2^2)}}},\sqrt{2}R\right)\label{bb4}\\
0,~\text{if}~\rho\in\left[\sqrt{2}R,\infty\right),\label{bb5}
\end{subnumcases}
\end{proposition}
where $\beta_0^*$ is any point within the regime of $\left[\frac{\rho}{R},1-\frac{\rho}{R}\right]$, $\beta_1^*$ is the unique solution to
$(q_1-q_2)\sqrt{\rho^2-\beta^2R^2}-q_1\sqrt{\rho^2-(1-\beta)^2R^2}=0$,
$\beta_2^*$ is unique solution to
$(q_1-q_2)(\beta{R}+\sqrt{\rho^2-\beta^2R^2})$ $-2q_1\sqrt{\rho^2-(1-\beta)^2R^2}=0$,
$\beta_3^*=1-\frac{\rho(q_1+q_2)}{R\sqrt{2(q_1^2+q_2^2)}}$, and $\beta_4^*$ is the unique solution to
$(1-\beta)(q_1-q_2)R+2q_2\sqrt{\rho^2-(1+\beta)^2R^2}-(q_1+q_2)\sqrt{\rho^2-(1-\beta)^2R^2}=0$.


From Proposition 8, the optimal $\beta^*$ not only depends on the target SNR $\gamma_{th}$ (which is reflected by $\rho$) but also the average traffic load $\mu$ (which is reflected by $q_j$).
If $\rho<R/2$, the optimal $\beta^*=[\rho/R,1-\rho/R]$ is the same as that of the 1D network in Proposition~5.
If $\rho\geq{R}/2$, the optimal $\beta^*$ in \eqref{bb1}-\eqref{bb5} depends on $\mu$ in the 2D network, which is a sharp contrast to $\beta^*=1-\rho/R$ regardless of $\mu$ for all $\rho\in[R/2,R]$ in the 1D network as shown in Proposition~5.
To provide more insight on how $\beta^*$ changes with $\rho$ (or $\gamma_{th}$) and $\mu$,  we further discuss the asymptotic results for the optimal $\beta^*$
for sufficiently low traffic load (as $\mu\rightarrow{0}$) and sufficiently high traffic load ($\mu\rightarrow\infty$), respectively.
\begin{corollary}
As $\mu\rightarrow{0}$, we have
\begin{subnumcases}{\beta^*=\nonumber}
\left[\frac{\rho}{R},1-\frac{\rho}{R}\right],~\text{if}~\rho\in\left[0,\frac{R}{2}\right)\label{bbb0}\\
\frac{1}{2},~~~~~~~~~~~~~~\text{if}~\rho\in\left[\frac{R}{2},\frac{\sqrt{2}R}{2}\right)\label{bbb1}\\
1-\frac{\rho}{\sqrt{2}R},~~~~~\text{if}~\rho\in\left[\frac{\sqrt{2}R}{2},\sqrt{2}R\right)\label{bbb3}\\
0,~~~~~~~~~~~~~~~\text{if}~\rho\in\left[\sqrt{2}R,\infty\right).
\end{subnumcases}
As $\mu\rightarrow{\infty}$, we have
\begin{subnumcases}{\beta^*=\nonumber}
\left[\frac{\rho}{R},1-\frac{\rho}{R}\right],~\text{if}~\rho\in\left[0,\frac{R}{2}\right)\label{bbbb0}\\
1-\frac{\rho}{R},~~~~~~~~\text{if}~\rho\in\left[\frac{R}{2},R\right)\label{bbb2}\\
0,~~~~~~~~~~~~~~\text{if}~\rho\in\left[R,\infty\right).\label{bbb4}
\end{subnumcases}
\end{corollary}
\begin{IEEEproof}
As $\mu\rightarrow{0}$, we have $q_1\rightarrow{1/4}$ and $q_2\rightarrow{0}$ according to Proposition 6. By substituting $q_1$ and $q_2$ into \eqref{bb1}, \eqref{bb3} and \eqref{bb4}, we further obtain $\beta^*=\beta_1^*=\frac{1}{2}$ for $\rho\in[R/2,\sqrt{2}R/2)$ and  $\beta^*=\beta_3^*=\beta_4^*=1-\frac{\rho}{\sqrt{2}R}$ for $\rho\in[\sqrt{2}R/2,\sqrt{2}R)$. We thus have $\beta_1^*\geq\beta_3^*$, where the equality holds if $\rho=\sqrt{2}R/2$. Moreover, we have $\beta^*=0$ for $\rho\in[\sqrt{2}R,\infty]$ as in \eqref{bb5}. Note, we do not have $\beta_2^*$ in \eqref{bb2} since  the lower and upper bounds of $\rho$ overlap with each other.

As $\mu\rightarrow{\infty}$, we have $q_1\approx{q}_2\rightarrow{1/16}$ according to Proposition 6. By substituting $q_1$ and $q_2$ into \eqref{bb1}, \eqref{bb2}, \eqref{bb3} and \eqref{bb4}, we have
$\beta^*=\beta_1^*=\beta_2^*=\beta_3^*=1-\frac{\rho}{R}$ for $\rho\in[0,R]$
and $\beta^*=\beta_4^*=0$ for $\rho\in[R,\sqrt{2}R]$. Furthermore, we have  $\beta^*=0$ for $\rho\in[\sqrt{2}R,\infty)$ as in \eqref{bb5}.
\end{IEEEproof}

\begin{figure}[t!]
    \begin{center}
        \includegraphics[width=0.9\columnwidth]{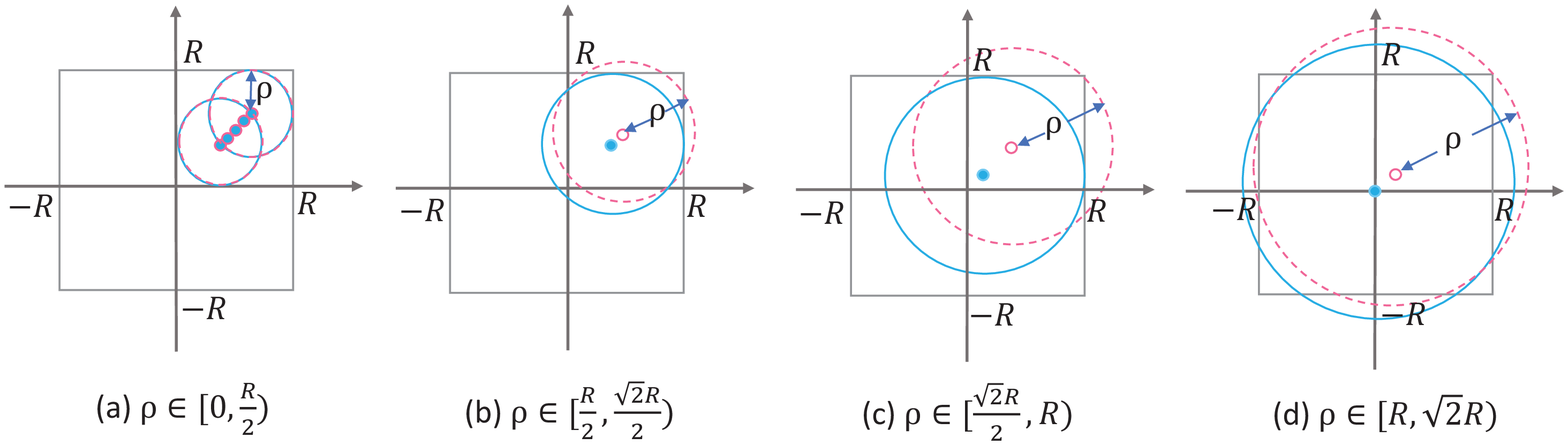}
        \caption{Graphical illustration of the asymptotically optimal UAV displacement in Corollary 5 given $k_1>\max(k_2,k_3,k_4)$ or $U=U_1$. The pink dashed circle and blue solid circle are the coverage regions for  $\mu\rightarrow{0}$ and $\mu\rightarrow{\infty}$, respectively. The pink hollowed dot and blue solid dot are the optimal UAV displacement locations for  $\mu\rightarrow{0}$ and $\mu\rightarrow{\infty}$, respectively.}
    \end{center}
\end{figure}

We now assume $k_1>\max(k_2,k_3,k_4)$ and give more intuitive explanations of Corollary 5. Basically, $\mu\rightarrow0$ tells the typical user is likely to be the only user in the cell and we thus decide the aggressive UAV deployment to cover $S_1$ as much as possible, while $\mu\rightarrow\infty$ tells that the users are more evenly distributed in all sectors and we thus decide the conservative UAV deployment that
covers the other sectors as well. More specifically, we have the following  discussions.
\begin{itemize}
\item If $\rho\in[0,{R}/{2}]$, the UAV's coverage region is far from enough to cover the whole $S_1$ as shown in Fig.~3(a).
    In this case, the optimal displacement location of the UAV is any point along the displacement direction that guarantees the circular coverage region of the UAV to be  within $S_1$.
    The corresponding results in \eqref{bbb0} and \eqref{bbbb0} are the same for any $\mu$, which is  consistent with Proposition 5 in  the 1D case.
\item  If $\rho\in[R/2,\sqrt{2}R/2)$, the UAV's coverage region is larger but still not enough to cover the whole $S_1$ including its corners  as shown in Fig.~3(b). When the traffic load is sufficiently low (as $\mu\rightarrow{0}$),
    given $k\geq{1}$ and $k_1>\max(k_2,k_3,k_4)$,
    it is very likely that $S_1$ is the only sector that has a user. The best strategy for the UAV is to move to the center of $S_1$, i.e., $\beta^*=1/2$ as in \eqref{bbb1} or the pink hollowed dot in Fig. 3(b).
    When the traffic load is sufficiently high (as $\mu\rightarrow{\infty}$), each sector has a  similar number of users though $k_1$ is still the largest.
    In this case, the UAV focuses more on serving users in $S_1$ while also serving the users in other sectors as much as it could.
    With $\beta^*=1-\rho/R$ in \eqref{bbb2} or the blue solid dot in Fig. 3(b), the UAV maximizes its coverage area in $S_1$ and avoids wasting any coverage outside the cell.
\item If $\rho\in[\sqrt{2}R/2,R)$, the coverage region of the UAV is larger than $S_1$ as shown in Fig.~3(c). As $\mu\rightarrow{0}$, the UAV will still focus on covering all points in $S_1$ and mildly cover other sectors by choosing $\beta^*=1-\frac{\rho}{\sqrt{2}R}$ as in \eqref{bbb3} or the pink hollowed dot in Fig. 3(c). It unnecessarily covers some points outside $S_1$. As $\mu\rightarrow{\infty}$, the other sectors are also important and the UAV will not cover any point outside $S_1$ by choosing $\beta^*=1-\frac{\rho}{R}$ in \eqref{bbb2} or the blue solid dot in Fig. 3(c).
\item  If $\rho\in[R,\sqrt{2}R)$, the UAV's coverage region is just not enough  to cover the whole square cell as shown in Fig.~3(d). As $\mu\rightarrow{0}$, similar to the previous case, the optimal displacement is $\beta^*=1-\frac{\rho}{\sqrt{2}R}$ in \eqref{bbb3} or the pink hollowed dot in Fig. 3(d). As $\mu\rightarrow{\infty}$, we have $\beta^*=0$ in \eqref{bbb4} and the blue solid dot in Fig. 3(d).
    Finally, as $\rho$ increases to be greater than $\sqrt{2}R$, the UAV can cover any point in the square cell and does not need to adapt to the user realizations.
\end{itemize}

\section{Simulation Results}

In this section, we evaluate the proposed adaptive UAV deployment scheme
for different performance objectives in the 1D and 2D networks.
We consider the downlink case and the UAV is operating with transmit power of $P_t=10$ mW at the height of $h=100$~m. The target cell for the 1D network is a line interval with the length of $R=1000$~m and that for the 2D network is a square region with the width of $2R=2000$~m.
The communication bandwidth is $1$ MHz and the carrier frequency is $5.8$~GHz.
We consider the noise power spectral density is $-170$~dBm$/$Hz and thus the receiver noise power is $\sigma^2=-110$~dBm.
The corresponding wavelength of the carrier frequency is $\nu=0.05$~m, so that the reference channel power at $d=1$~m is $\theta=-47$ dB. We use $10^7$ number of realizations to generate all simulation results in this section.

In the previous sections, we adopt the simple and practical majority-vote based adaptive scheme for tractable analysis, where the UAV does not need to adapt with a different $\beta^*$ in each realization.
We now further discuss some other adaptive/non-adaptive schemes based on different side information on the user number and even their locations, which are defined as:
\begin{itemize}
\item Adaptive deployment with perfect user location knowledge (Perfect knowledge adaptive scheme): the UAV has the perfect knowledge of the total number of  MUs and their exact locations in each realization. The UAV updates its optimal displacement  in each realization to maximize the average throughput or successful transmission events of all MUs. Under this scheme, the optimal displacement factor $\beta^*(X_m)$ is a function of user coordinates $X_m$ in each realization.
\item Adaptive deployment with exact user number per sector (Exact user number adaptive scheme): the UAV only knows the number of users $k_i$ for each sector (but not the user exact locations therein) in each realization. It updates its displacement location in each realization to maximize the expected outcome of the average throughput or successful transmission probability of all MUs. Under this scheme, the optimal displacement factor $\beta^*(k_i)$ is a function of user numbers $k_i$ in each realization. For the 1D scenario, the analytical result of $\beta^*(k_i)$ for average throughput maximization is given in Appendix A and that for success probability maximization is the same as Proposition 5 as discussed in Section IV.
\item Non-adaptive benchmark scheme: the UAV keeps staying at the cell center due to the homogeneous user density. We always have $\beta^*=0$ in all realizations.
\end{itemize}

\begin{figure}[t!]
    \begin{center}
        \includegraphics[width=0.95\columnwidth]{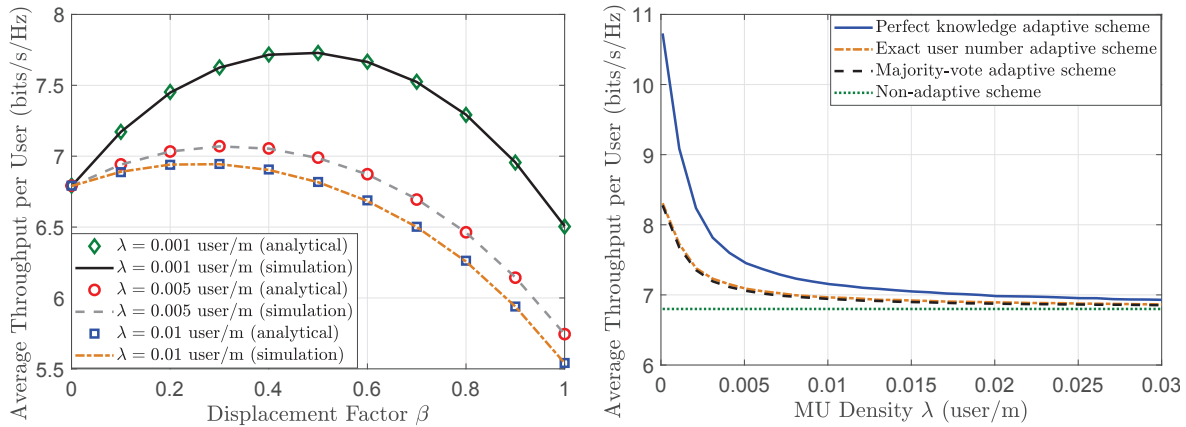}
        \caption{(a) Optimal $\beta^*$ that maximizes the average throughput in the 1D user network. (b) Comparison of the maximum average throughput $\mathbb{E}[C(\beta^*)]$ among various schemes.}
    \end{center}
\end{figure}

First, in Fig. 4 and Fig. 5, the performance is shown for the 1D user network case. In Fig.~4(a), the average throughput of MU$_0$ is observed to be concave in $\beta$, where the simulation results match well with that of Proposition 2. We see that the optimal $\beta^*$
decreases with $\lambda$.
From Fig.~4(b), we observe that, if the UAV is able to adapt to the perfect knowledge of  user number and  locations, the performance gain is significant, though this is difficult to realize in practice and its advantage decreases with $\lambda$.
Compared to the scheme with exact user number per sector known,
the majority-vote scheme is easier to implement and also achieves a very close performance for
various user densities, explained as follows.
Intuitively, only when the user number is very asymmetric across all sectors in one realization, the exact user number adaptive scheme that adopts a different displacement factor customized for
this realization can obviously outperform the majority-vote scheme that adopts the same displacement factor across all realizations.
However, this extremely asymmetric event happens rarely under the HPPP setting as all sectors have same user density $\lambda$.
When we consider  the average throughput over the long run, this advantage is minor and thus the two schemes have very  close performance.
Furthermore, we observe that the three adaptive schemes greatly outperform the non-adaptive scheme, especially for  small user density.
When the traffic load is high,
all realizations tend to approach average sense and it is better to be non-adaptive to meet instantaneous traffic.

\begin{figure}[t!]
    \begin{center}
        \includegraphics[width=0.95\columnwidth]{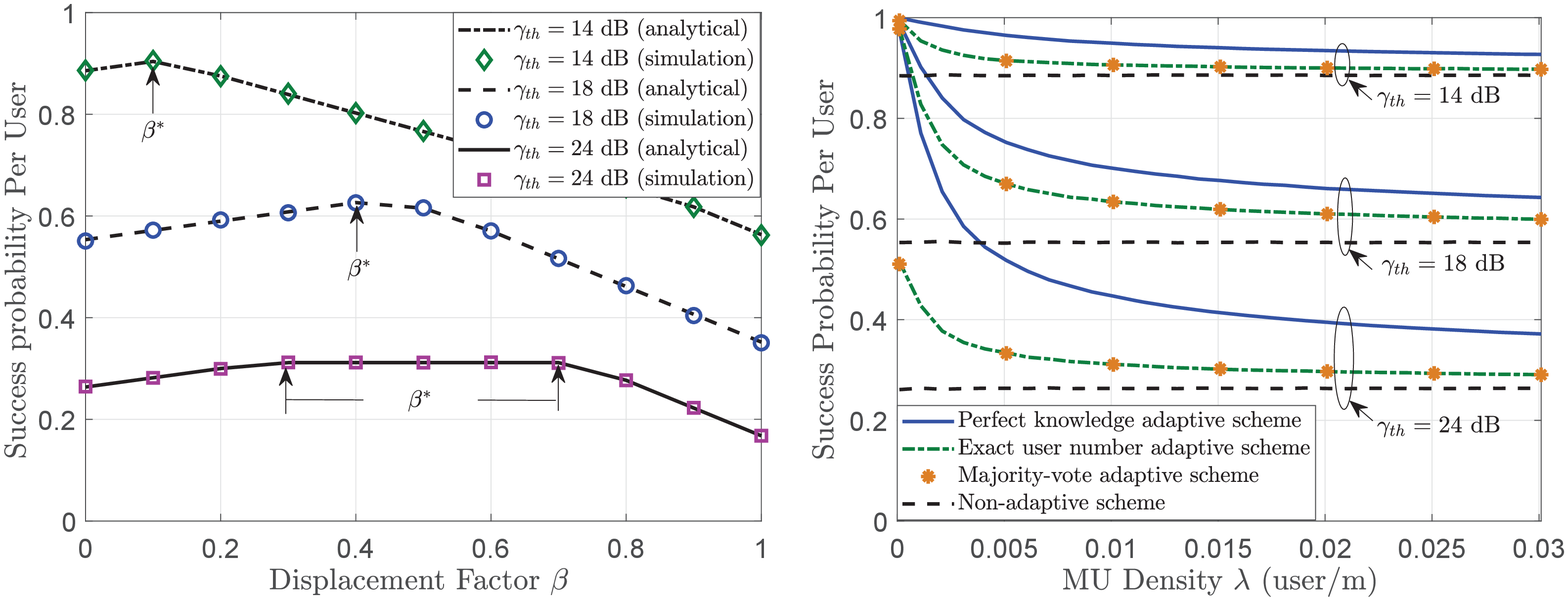}
        \caption{(a) Optimal $\beta^*$ that maximizes the success probability in the 1D user network. (b) Comparison of the maximum success probability $p(\beta^*)$ among various schemes.}
    \end{center}
\end{figure}

In Fig. 5(a), we  show that the success probability $p$ of the typical user is concave in  $\beta$ in the 1D network case, where the simulation results match the analytical results in Proposition 4.
In Fig. 5(b), we compare the four schemes in terms of success probability. Similar to Fig. 4(b), the maximum success probabilities of the three adaptive schemes greatly outperform that of the non-adaptive scheme for low traffic load and the performance gain decreases with the traffic load.
We further notice that the curve of the majority-vote scheme is aligned with that of the exact user number scheme.
Moreover, if the UAV knows the perfect number and exact locations of the MUs (though difficult in practice), the success probability can be further improved, where this improvement is more significant in the high target SNR regime. Intuitively, if the UAV has smaller coverage region, knowing exact MUs' locations helps better  pin-point the target MUs.

\begin{figure}[t!]
    \begin{center}
        \includegraphics[width=0.95\columnwidth]{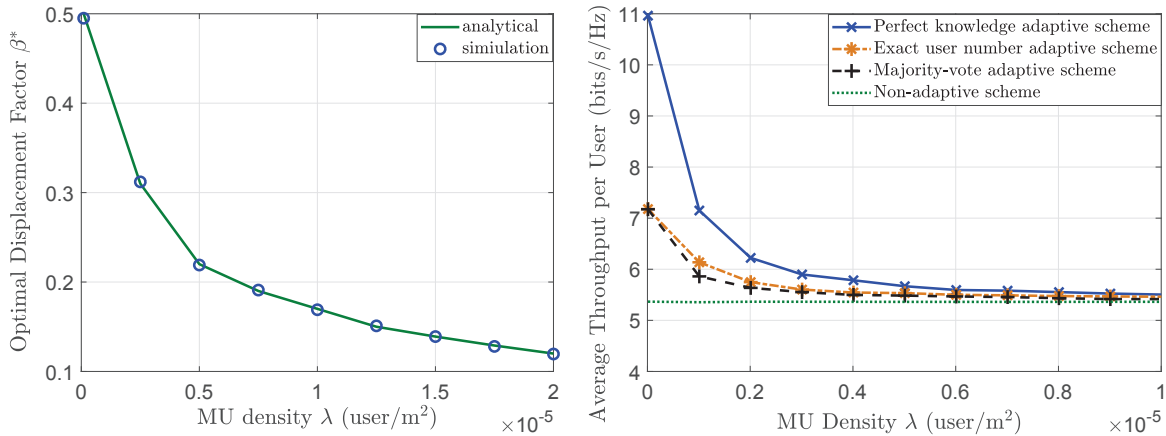}
        \caption{(a) Optimal $\beta^*$ that maximizes the average throughput in the 2D user network. (b) Comparison of the maximum average throughput $\mathbb{E}[C(\beta^*)]$ among various schemes.} 
    \end{center}
\end{figure}

Next, in Fig. 6 and Fig. 7, we show the performance in the 2D user network case. In Fig.~6(a), we show that the optimal displacement factor $\beta^*$ that maximizes the average throughput decreases with the MU density $\lambda$. The simulation result is consistent with the analytical result given in Proposition 7.
In Fig. 6(b),
we use exhaustive search method to find the optimal UAV deployment position (including both the optimal direction and optimal distance)
for the perfect knowledge adaptive scheme and the exact user number adaptive scheme, respectively.
The insight for the 2D network is similar to that of the 1D network except that the gap between the exact user number scheme and majority-vote scheme is slightly larger, which is due to the loss of optimality in the direction selection for the latter scheme  in the 2D network. Specifically, the two schemes are equivalent as $\lambda\rightarrow{0}$ when there is only one user (if any) in the cell.

\begin{figure}[t!]
    \begin{center}
        \includegraphics[width=0.95\columnwidth]{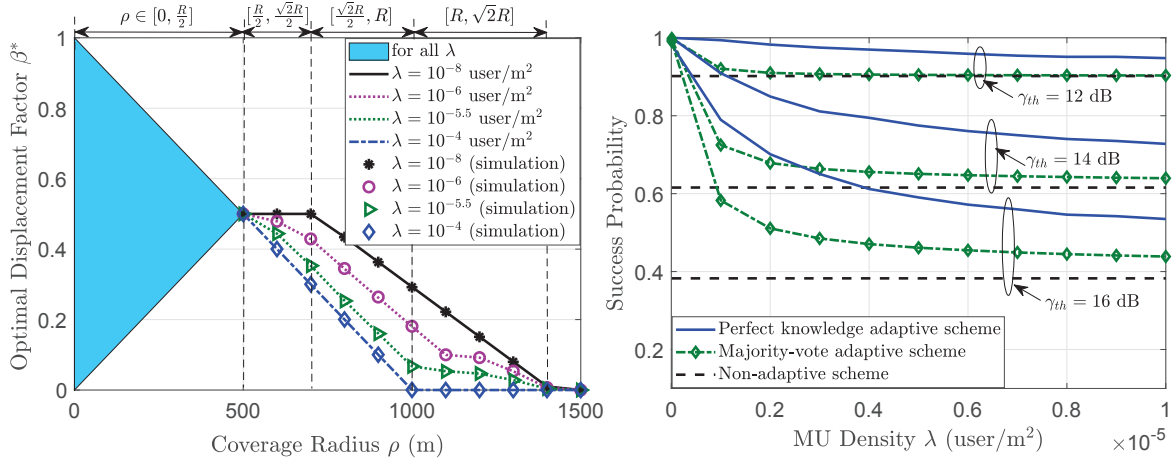}
        \caption{(a) Optimal $\beta^*$ that maximizes the success probability in the 2D user network. (b) Comparison of the maximum success probability $p(\beta^*)$ among various schemes.}
    \end{center}
\end{figure}

In Fig. 7(a), we plot the optimal $\beta^*$ that maximizes the success probability of MU$_0$ versus the UAV's coverage radius $\rho$. It matches well with the  analytical results in Proposition 8. If $\rho<R/2$, the optimal $\beta^*$ is flexible and can be any point  within the regime of $[\rho/R,1-\rho/R)$ as illustrated by the blue solid triangle, where this optimal regime shrinks with $\rho$.
If $\rho\in[R/2,\sqrt{2}R]$, the optimal $\beta^*$ is unique and generally decreases with increasing $\rho$. Intuitively,
the UAV  can move with a shorter distance when it has a wider coverage region.
Moreover, the optimal $\beta^*$ decreases as the user density $\lambda$ increases.
One can also check that the curves with $\lambda=10^{-8}$ and $\lambda=10^{-4}$ are consistent with the asymptotic discussions of $\mu\rightarrow{0}$ and $\mu\rightarrow\infty$ in Corollary 5.
In Fig.~7(b), we compare the maximum success probability for the majority-vote scheme with the perfect knowledge adaptive scheme and non-adaptive scheme under different SNR thresholds,\footnotemark~where the insight is similar to that of Fig. 5(b) in the 1D network case.
\footnotetext{We omit the exact user number scheme here due to the intractable complexity as a result of  the random MU locations in the exhaustive search for the optimal UAV location.}

\begin{figure}[t!]
    \begin{center}
        \includegraphics[width=0.55\columnwidth]{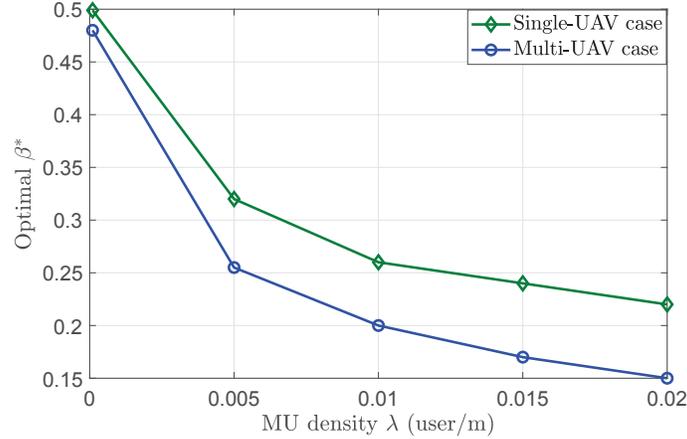}
        \caption{Comparison between the single-UAV and multi-UAV cases for 1D scenario.}
    \end{center}
\end{figure}

In Fig. 8, we extend the results to the multi-UAV case for 1D scenario.
We consider $2n+1$ UAV cells
by copying the single cell of $[-R,R]$ in Fig. 1 $n$ times on both sides  the line.
The target cell of UAV$_i$ is denoted by $[(i-1)R,(i+1)R]$, where $i$ is any integer within $[-n,n]$. Here we adopt $n=10$ for this simulation.  
The MUs follow HPPP with density of $\lambda$ over the whole 1D line, where the total user number inside each cell is
a Poisson random variable with the mean value of $\mu=2\lambda{R}$.
Similar to the single-UAV case, each UAV$_i$ also has three candidate stop points, i.e., $(iR-\beta{R},h)$, $(iR,h)$ and $(iR+\beta{R},h)$, and the UAV adapts its location following the similar majority-vote criteria as in (1).
We consider the worst case scenario of full frequency reuse among the cells, where all the UAVs are interfering with each other.
From Fig. 8 we observe that the optimal $\beta^*$ of the multi-UAV case is smaller than that of the single-UAV case for various user densities. Intuitively, the UAV should be more conservative in the adaptation process (i.e., not to move too close to each other) in order to avoid interfering with other cells.
Still, the insight is similar to the single-user case, where the optimal $\beta^*$ decreases with the user density $\lambda$.

\section{Conclusions}

In this paper, we proposed an adaptive UAV deployment scheme in a Poisson distributed 1D/2D random user network, where the UAV adapts its location according to the instantaneous traffic load in different sectors within its target cell.
We adopted a simple majority-vote rule to displace the UAV in the direction of the sector that has the highest number of users in each network realization. This scheme is applicable for the scenario when the exact user number/locations in each sector are difficult to obtain in practice. We designed the optimal displacement distance in the chosen sector to maximize the average throughput for the variable-rate application  and the success probability for the fixed-rate application, respectively.
For average throughput maximization, the optimal displacement distance decreases with the average traffic load. For success probability maximization, the optimal displacement distance
 does not necessarily decrease with the average traffic load but depends on the target SNR.
Extensive simulations show that the adaptive deployment scheme outperforms that of the non-adaptive scheme, especially for low traffic load.
In the future work, we are working towards generalizing the adaptive UAV deployment in a multi-antenna and/or multi-UAV scenario.

\section*{Appendix A: Derivation of Optimal $\beta^*$ Based on Exact User Number per Sector  for Average Throughput Maximization in 1D Network}
First, we derive the expected throughput of MU$_0$ in one realization with given known $k_1=K_1$ and $k_2=K_2$,
and the UAV location $U=U_j$ based on (1) as follows:
\begin{align}
&\mathbb{E}[C\big|k_1=K_1,k_2=K_2,U=U_j]\nonumber\\
&=\sum\limits_{i=1}^{2}\Pr\left[X_0\in{S_i}\big|k_1=K_1,k_2=K_2,U=U_j\right]\mathbb{E}\left[C\big|k_1=K_1,k_2=K_2,U=U_j,X_0\in{S_i}\right]\nonumber,
\end{align}
where
$\Pr\left[X_0\in{S_i}\big|k_1=K_1,k_2=K_2,U=U_j\right]=\frac{K_i}{K},~\text{for}~i=1,2$ and the conditional throughput of  $\mathbb{E}\left[C\big|k_1=K_1,k_2=K_2,U=U_j,X_0\in{S_i}\right]$ equals $\omega_{i,j}$
given in \eqref{omegaij}. Due to the symmetric feature, we further have $\omega_{1,0}=\omega_{2,0}=\omega_0$, $\omega_{1,1}=\omega_{2,2}=\omega_1$ and $\omega_{1,2}=\omega_{2,1}=\omega_2$. The expected throughput is thus rewritten as
\begin{subnumcases}
{\mathbb{E}\left[C\big|k_1=K_1,k_2=K_2,U=U_j\right]=}
\omega_0=\vartheta,~~~~~~~~~~~~~~~~~~~~~~~~~~~~~~~~~~~~~~~~~~\text{if $U=U_0$}\nonumber\\
\frac{K_1}{K}\omega_1+\frac{K_2}{K}\omega_2=\frac{K_1}{K}(\zeta+\kappa)+\frac{K_2}{K}(\xi-\kappa),~\text{if $U=U_1$}\nonumber\\
\frac{K_1}{K}\omega_2+\frac{K_2}{K}\omega_1=\frac{K_2}{K}(\zeta+\kappa)+\frac{K_1}{K}(\xi-\kappa),~\text{if $U=U_2$,}\nonumber 
\end{subnumcases}
where $K=K_1+K_2$.
Similar to Corollary 2 and Proposition 3,
we can prove that the expected throughput $\mathbb{E}\left[C\big|k_1=K_1,k_2=K_2,U=U_j\right]$ for $j\neq0$ is strictly concave in $\beta$ for $\beta\in[0,1]$
and the optimal $\beta^*(k_1,k_2)$ in this realization is the solution to
$K_j\varrho_1
-K_i\varrho_2=0~\text{for}~j=1,2~\text{and}~i\neq{j}$.

\section*{Appendix B: Proof of Proposition 6}
We rewrite \eqref{qjj}  as
\begin{align}
q_j
&=
{\Pr\left[k_j>\max\limits_{i\neq{j}}(k_i),W_0\in{S_1},k\geq{1}\right]}/{\Pr\left(k\geq{1}\right)},\label{k1}
\end{align}
where $\Pr\left[k\geq{1}\right]=1-\exp(-\mu) $ with $\mu=4{R^2}\lambda$.

\subsection*{A. Derivations of $q_j$}
We first derive the joint probability in the numerator of \eqref{k1} for $j=1$. Since $W_0\in{S_1}$ indicates that at least one user is inside the sector $S_1$, we can rewrite  $k\geq{1}$ as $k_1\geq{1}$ and $k_i\geq{0}$ $\forall~i\neq{1}$ and thus have
\begin{align}
&\Pr\left[k_1>\max\limits_{i\neq{1}}(k_i),W_0\in{S_1},k\geq{1}\right]
=\Pr\left[k_1>\max\limits_{i\neq{1}}(k_i),W_0\in{S_1},k_1\geq{1},k_{i}\geq{0} ~\forall{i}\neq{1}\right].\nonumber
\end{align}
Note that we cannot remove the event of $W_0\in{S_1}$ from the above joint events. Though we can deduce $k_1\geq{1}$ from $W_0\in{S_1}$, the converse is not true. 
We adopt convolution to derive this joint probability.
Let $k_1=t_1$ and $k_i=t_1-t_i$ $\forall$ $i\neq{1}$. Since $k_1\geq{1}$ and $k_{i}\geq{0}$ $\forall$ $i\neq{1}$, we have $t_1\geq{\max\limits_{i\neq{1}}(1,t_i)}$. Moreover, since $k_1>\max\limits_{i\neq{1}}(k_i)$ and $k_i\geq0$, we have $t_i\geq{1}$ $\forall~i$. By averaging over $k_1$ and all possible $k_i$ for $i\neq{1}$, the joint probability is
\begin{align}
&\Pr\left[k_1>\max\limits_{i\neq{1}}(k_i),W_0\in{S_1},k_1\geq{1},k_{i}\geq0~\forall{i}\neq1\right]\nonumber\\
&=\sum_{t_M=1}^{\infty}\cdots\sum_{t_2=1}^{\infty}\sum_{t_1=\max\limits(t_2,\cdots,t_M)}^{\infty}
\Pr\left[W_0\in{S_1}\big|k_1=t_1,k_2=t_1-t_2,\cdots,k_M=t_1-t_M\right]\nonumber\\
&~~~\times\Pr\left[k_1=t_1,k_2=t_1-t_2,\cdots,k_M=t_1-t_M\right]\label{eq1}.
\end{align}
As $k\sim{\text{Poi}(\mu)}$ and $k_j\sim{\text{Poi}(\mu/M)}$ are independent, we decompose the joint PDF into $M$ independent Poisson PDFs.
\begin{align}
&\Pr\left[k_1=t_1,k_2=t_1-t_2,\cdots,k_M=t_1-t_M\right]
=\Pr\left[k_1=t_1\right]\prod_{i=2}^{M}\Pr\left[k_i=t_1-t_i\right]\nonumber\\
&=\left[e^{-\frac{\mu}{M}}\frac{\left(\frac{\mu}{M}\right)^{t_1}}{t_1!}\right]
\left[e^{-\frac{\mu}{M}}\frac{\left(\frac{\mu}{M}\right)^{t_1-t_2}}{(t_1-t_2)!}\right]\cdots
\left[{e}^{-\frac{\mu}{M}}\frac{\left(\frac{\mu}{M}\right)^{t_1-t_M}}{(t_1-t_M)!}\right]
=e^{-\mu}\frac{\left(\frac{\mu}{M}\right)^{Mt_1-\sum_{i=2}^{M}t_i}}{t_1!\prod_{i=2}^{M}(t_1-t_i)!}.\label{eq2}
\end{align}
Conditioned on the fixed values of each $k_i$, the probability that the typical user is selected from the first sector, i.e., $W_0\in{S_1}$, is given by $\frac{k_1}{\sum_{i=1}^Mk_i}$. We thus have
\begin{align}
\Pr\left[W_0\in{S_1}\big|k_1=t_1,k_2=t_1-t_2,\cdots,k_M=t_1-t_M\right]
=\frac{t_1}{Mt_1-\sum_{i=2}^{M}t_i}.\label{eq3}
\end{align}
By substituting \eqref{eq2} and \eqref{eq3} into \eqref{eq1}
and \eqref{k1},
we obtain  $q_1$ in Proposition 6.

Next, we derive the joint probability in the numerator of \eqref{k1} for $j\geq{2}$.
We can deduce $k_1\geq{1}$ from $W_0\in{S_1}$ and further deduce $k_j\geq{2}$ since $k_j>k_1$. The event of $k\geq{1}$ is thus equivalent to $k_1\geq{1}$, $k_j\geq{2}$ and $k_i\geq{0}$ $\forall{i}\neq{1,j}$.
The joint probability is thus given by
\begin{align}
&\Pr\left[k_j>\max\limits_{i\neq{j}}(k_i),W_0\in{S_1},k\geq{1}\right]\nonumber\\
&=\sum_{t_M=1}^{\infty}\cdots\sum_{t_2=1}^{\infty}
\sum_{t_1=\max(t_2+1,t_3,\cdots,t_M)}^{\infty}\Pr\left[k_j=t_1\right]
\Pr\left[k_1=t_1-t_2\right]\prod_{i\neq{1,j}}\Pr\left[k_i=t_1-t_i\right]\nonumber
\end{align}
\begin{align}
&~~~\times\Pr\left[W_0\in{S_1}\big|k_j=t_1,k_1=t_1-t_2,k_{i(\forall{i}\neq{1,j})}=t_1-t_i\right]\nonumber\\
&=e^{-\mu}\sum_{t_M=1}^{\infty}\cdots\sum_{t_2=1}^{\infty}
\sum_{t_1=\max(t_2+1,t_3,\cdots,t_L)}^{\infty}
\frac{\left(\frac{\mu}{M}\right)^{Mt_1-\sum_{i=2}^{M}t_i}}{t_1!\prod_{i=2}^{M}(t_1-t_i)!}
\frac{t_1-t_2}{Mt_1-\sum_{i=2}^Mt_i}.\label{eq5}
\end{align}
By substituting \eqref{eq5} into \eqref{k1}, we obtain the second case of $q_j$ in Proposition 6.

Third, we derive the joint probability in the numerator of \eqref{k1} for $j=0$.
In this case, the UAV stays at the origin, i.e., $U=U_0$. The joint probability is given by
\begin{align}
q_0
&=\Pr\left(W_0\in{S_1}\big|k\geq{1}\right)
-\sum_{i=1}^{M}\Pr\left[U=U_i,W_0\in{S_1}\big|k\geq{1}\right].\label{q0}
\end{align}
As $k\sim{\text{Poi}(\mu)}$ and the $M$ sectors are equally partitioned,
we have $k_1\sim\text{Binom}(k,\frac{1}{M})$.
As the typical user is chosen from these MUs, we thus have
$\Pr\left[W_0\in{S_1}\big|k\geq{1}\right]=\frac{1}{M}$.
Substituting it into \eqref{q0}, we obtain the third case of $q_j$ in Proposition 6.

\subsection*{B. Asymptotic Analysis}
Conditioned on $k\geq{1}$, the conditional PMF  is given by \eqref{ConK0} and \eqref{ConK1}.
As $\mu\rightarrow{0}$, we have $e^{-\mu}=1-\mu$.
For $j=1$, we have
\begin{align}
\lim\limits_{\mu\rightarrow{0}}q_1
&=\lim\limits_{\mu\rightarrow{0}}\Pr\left[k=1\big|k\geq{1}\right]\Pr\left[k_1=1,k_2=0,\cdots,k_L=0\big|k={1}\right]\nonumber\\
&~~~\times\Pr\left[W_0\in{S_1}\big|k_1=1,k_2=0,\cdots,k_M=0\right]
=1\times{\frac{1}{M}}\times{1}=\frac{1}{M}.
\end{align}
For $j\geq{2}$, we have
\begin{align}
\lim\limits_{\mu\rightarrow{0}}q_j
&=\lim\limits_{\mu\rightarrow{0}}\Pr\left[k=1\big|k\geq{1}\right]
\Pr\left[k_j>\max\limits_{i\neq{j}}(k_i),W_0\in{S_1}\big|k=1\right]=1\times{0}=0.
\end{align}
And for $j=0$, we have $\lim\limits_{\mu\rightarrow{0}}q_0=0$. Similarly, we can prove that $\lim\limits_{\mu\rightarrow{\infty}}q_j=1/M^2$ for $j\neq{0}$ and $\lim\limits_{\mu\rightarrow{\infty}}q_0=0$.


\bibliographystyle{plainnat}

\end{document}